\documentclass[amsmath,amssymb,11pt]{article}
\usepackage{jheppub2}
\pdfoutput=1
\usepackage{graphicx,epsfig}
\usepackage{float}
\usepackage{subfig}
\usepackage{subfloat}
\usepackage[utf8]{inputenc}
\usepackage[T1]{fontenc}

\usepackage{hyperref}
\usepackage{subcaption}
\usepackage{color}
\usepackage{bbm}
\usepackage{indentfirst}
\usepackage{ amssymb }
\usepackage{amsmath}

\usepackage{mathtools}
\usepackage{amsfonts}
\usepackage{physics}
\usepackage{tikz}
\usepackage{comment}
\usepackage{soul}

\setstcolor{red}

\newcommand{\beq}{\begin{equation}}

\newcommand{\eeq}{\end{equation}}

 \newcommand{\be}{\begin{equation}}
 \newcommand{\ee}{\end{equation}}
 \newcommand{\bea}{\begin{eqnarray}}
 \newcommand{\eea}{\end{eqnarray}}

\usepackage{tikz}
\usetikzlibrary{positioning}
\usetikzlibrary{intersections}
\usetikzlibrary{fadings} 
\usetikzlibrary{arrows.meta} 
\usetikzlibrary{arrows}

\tikzfading[name=fade out,
inner color=transparent!0,
outer color=transparent!100]

\definecolor{cherryblossompink}{rgb}{1.0, 0.72, 0.77}
\definecolor{lightblue}{rgb}{0.68, 0.85, 0.9}

\usetikzlibrary{decorations.pathmorphing}
\usetikzlibrary{decorations.pathreplacing,decorations.markings}

\usetikzlibrary{backgrounds,automata}

\title{Untangling Selberg from the Wilson spool:\\ 1-loop determinants and trace formulae in (A)dS$_{3}$} %

\author[a]{Samuel Haupfear,}
\author[b]{Victoria Martin,}
\author[c]{Andrew Svesko,}
\author[a]{and Claire Zukowski}

\affiliation[a]{Department of Physics and Astronomy, University of Minnesota Duluth, Duluth, MN 55812, USA}
\affiliation[b]{Department of Physics, University of North Florida, Jacksonville, FL 32224, USA}
\affiliation[c]{Department of Mathematics, King’s College London, Strand, London, WC2R 2LS, United Kingdom}

\emailAdd{haupf004@umn.edu}
\emailAdd{victoria.martin@unf.edu}
\emailAdd{andrew.svesko@kcl.ac.uk}
\emailAdd{czukowsk@d.umn.edu}

\abstract{Leading quantum effects in perturbative quantum gravity are captured by functional determinants of kinetic operators. We study such 1-loop determinants in three-dimensional Euclidean (anti-) de Sitter gravity evaluated using two seemingly disparate tools, the Selberg zeta function and the Wilson spool. For the Euclidean BTZ black hole, we demonstrate the Wilson spool for massive bosons of arbitrary spin directly equates to a representation-theoretic version of the Selberg zeta function.
In the case of Euclidean de Sitter, we show a new trace formula associated with the Fredholm determinant for the scalar Laplacian on the three-sphere reproduces the Wilson spool. Generalizing the trace formula, we comment on how to extend this Wilson spool construction to lens space quotients and higher-dimensional spheres. 
}

\begin{document}

\maketitle

\section{Introduction} \label{sec:intro} 

\noindent In the path integral formulation of quantum gravity, the fundamental object of interest is the gravitational partition function \cite{Gibbons:1976ue}
\beq \mathcal{Z}=\int Dg D\phi e^{-I_{E}[g,\phi]}\;.\label{eq:gravpartgen}\eeq
The integration measure indicates a sum over all dynamical (Euclidean) metrics and matter field configurations $\phi$ in a theory characterized by its Euclidean action, $I_{E}[g,\phi]$.
In nearly all scenarios, the partition function (\ref{eq:gravpartgen}) is impossible to compute exactly.
In practice, $\mathcal{Z}$ is evaluated perturbatively by expanding $I_{E}$ about solutions $\{g_{c},\phi_{c}\}$ to the classical equations of motion.
Let $\{h,\psi\}$ denote small fluctuations about these saddles. For each saddle-point, the action expands as $I_{E}[g,\phi]=I^{\text{on-shell}}_{E}[g_{c},\phi_{c}]+I^{(2)}_{E}[h,\psi]+...$, where 
 $I^{(2)}_{E}[h,\psi]$ is quadratic in field fluctuations and is the leading (1-loop) perturbative quantum correction to $I_{E}[g,\phi]$.  Heuristically, the partition function (\ref{eq:gravpartgen}) may be cast as
\beq
\begin{split}
\mathcal{Z}&\approx \sum_{\{g_{c},\phi_{c}\}} e^{-I_{E}^{\text{on-shell}}[g_{c},\phi_{c}]}\int Dh D\psi e^{-I_{E}^{(2)}[h,\psi]+...}\equiv\sum_{\{g_{c},\phi_{c}\}}Z^{(0)}[g_{c},\phi_{c}]Z^{(1)}[g_{c},\phi_{c}]+...\;\;.
\end{split}
\eeq
Famously, the entropy-area relation of black holes \cite{Bekenstein:1973ur,Hawking:1974sw} and cosmological horizons \cite{Gibbons:1977mu} can be derived from the tree-level partition function $Z^{(0)}$. Meanwhile, leading quantum corrections to horizon thermodynamics (e.g., \cite{Sen:2008vm,Banerjee:2010qc,Sen:2012kpz,Anninos:2020hfj}) are encoded in the 1-loop partition function, $Z^{(1)}$. 

Generally, evaluating the 1-loop partition function amounts to computing functional determinants of differential kinetic operators. This is because the quadratic action schematically is of the form $I^{(2)}[h,\psi]=\int d^{d}x\sqrt{g}\xi\nabla_{\xi}^{2}\xi$, where $\xi$ collectively denotes the field fluctuations, and $\nabla_{\xi}^{2}$ is a differential kinetic operator evaluated in the classical background. 
Concretely, for a free (real) scalar field $\phi$ of mass $m$, the 1-loop partition function is
\begin{equation}\label{Eq:Partitionfunction}
    Z^{(1)} = \int D\phi e^{-\frac{1}{2}\int_{\mathcal{M}} d^{d}x\sqrt{g}\phi \left(-\nabla^2+m^2\right)\phi}=\text{det}\left(-\nabla^2+m^2\right)^{-1/2}\;,
\end{equation}
where metric fluctuations are ignored, $\mathcal{M}$ is the Euclidean section of a $d$-dimensional manifold, and $\nabla^{2}=g^{\mu\nu}\nabla_{\mu}\nabla_{\nu}$ is the scalar Laplacian with respect to background metric $g$. 
There are multiple techniques for evaluating such 1-loop determinants, such as heat kernel methods~\cite{Giombi:2008vd,David:2009xg,Gopakumar:2011qs} and quasinormal mode~\cite{Denef:2009kn,Denef:2009yy,Datta:2011za,Castro:2017mfj} methods, each with (dis)advantages and physical insights. 

In this article, we connect two previously distinct methods for computing 1-loop functional determinants in three-dimensional perturbative quantum gravity. Specifically, we show, for arbitrary spin bosons, the Selberg zeta function \cite{selberg1956harmonic} and the Wilson spool \cite{Castro:2023dxp,Castro:2023bvo} for Euclidean anti-de Sitter (AdS) space are essentially equivalent. Meanwhile, though there is no known analog of the Selberg zeta function for Euclidean dS$_{3}$, i.e., the 3-sphere $S^{3}$, we highlight mathematical similarities the $S^{3}$ Wilson spool shares with Selberg zeta functions of three-dimensional hyperbolic quotients $\mathbb{H}^{3}/\mathbb{Z}$. Building off of these observations,  we provide a novel derivation of the Wilson spool for a massive real scalar field in Euclidean de Sitter (dS) space by constructing a new trace formula for the Fredholm determinant. Notably, we straightforwardly extend our analysis to lens space quotients of $S^{3}$, thus offering a benchmark for a direct construction of the Wilson spool on lens spaces. 

To appreciate our findings, let us briefly summarize the essential elements.

\vspace{3mm}

\noindent \textbf{Selberg zeta function.} In Euclidean AdS$_{3}$ gravity, it is of interest to study thermal geometries, e.g., thermal AdS$_{3}$ or the Euclidean Ba{\~n}ados-Teitelboim-Zanelli (BTZ) black hole \cite{Banados:1992wn,Banados:1992gq}; both are examples of hyperbolic quotients $\mathbb{H}^{3}/\Gamma$ for a discrete subgroup $\Gamma$ of $\text{SL}(2,\mathbb{C})$. For such backgrounds, there exists the (Patterson-) Selberg zeta function \cite{Patterson1989TheSZ}
\begin{equation}\label{Eq:Selberg3D}
    \zeta_{\mathbb{H}^3/\Gamma}(\mathfrak{s})=\prod_{k_1,k_2=0}^{\infty}\left[1-\alpha_{1}^{k_1}\alpha_{2}^{k_2}e^{-(\mathfrak{s}+k_1+k_2)\ell(p)}\right]\;.
\end{equation}
Here, $\alpha_i$ are the eigenvalues of group element $O(3)\subset \Gamma$, and $\ell(p)$ is the length of the `prime geodesic' $p$.\footnote{Historically, the Selberg zeta function was originally defined for two-dimensional quotients $\mathbb{H}^{2}/\Gamma$ for $\Gamma\subset \text{SL}(2,\mathbb{R})$ \cite{selberg1956harmonic}, as an analog to the Riemann zeta function $\zeta(\mathfrak{s})=\prod_{p\in\mathbb{P}}(1-p^{-\mathfrak{s}})^{-1}$ for prime numbers $\mathbb{P}$. Specifically, $\zeta_{\mathbb{H}^{2}/\Gamma}=\prod_{p}\prod_{k=0}^{\infty}(1-\ell(p)^{-(\mathfrak{s}+k)})$, where $p$ runs over conjugacy classes of `prime geodesics', i.e.,  primitive closed geodesics (closed geodesics that trace out their image exactly once). The form (\ref{Eq:Selberg3D}) is a special case of the Patterson-Selberg zeta function defined for general Kleinian groups \cite{Patterson1989TheSZ}.\label{fn:primegeodes}}
Pertinently, the Selberg zeta function is directly related to the 1-loop partition function~\cite{doi:10.1142/S0217751X03015660,Perry2003SELBERGZF,Williams:2012ms,Diaz:2008iv,Aros:2009pg,Williams:2015azf,Aros:2016opw,Keeler:2018lza,Keeler:2019wsx,Martin:2019flv,Martin:2022duk,Bagchi:2023ilg}. For example, the (regularized) 1-loop partition function for a massive scalar field on $\mathbb{H}^{3}/\mathbb{Z}$ is
\beq\label{eq:selr} Z^{(1)}(\Delta)=\frac{1}{\zeta_{\mathbb{H}^{3}/\mathbb{Z}}(\Delta)}\;,\eeq
where $\Delta$ is a specific function of the mass of the scalar field. 

The Selberg zeta function method for evaluating 1-loop determinants has multiple advantages. Firstly, it provides a computationally simple way of evaluating 1-loop determinants; the only necessary inputs are the eigenvalues $\alpha_{i}$ and geodesic length, both of which are well-known for arbitrary spin fields on $\mathbb{H}^{3}/\mathbb{Z}$, i.e., thermal AdS$_{3}$ and the Euclidean BTZ black hole. Further, since the Selberg zeta function is known for $n$-dimensional hyperbolic quotients, $\mathbb{H}^{n}/\Gamma$, it can readily be used to compute functional determinants in higher- or lower-dimensional gravity, e.g., \cite{Martin:2019flv}.  Moreover, if any two of (i) zeros of the Selberg zeta function, (ii) thermal Matsubara frequencies, or (iii) Lorentzian quasinormal modes (QNMs) are given, the third can
be determined \cite{Keeler:2018lza,Keeler:2019wsx}. This finding was used to directly construct Lorentzian QNMs for various geometries~\cite{Martin:2022duk,Bagchi:2023ilg}. Finally, the Selberg zeta function has generalizations beyond hyperbolic quotients, such as warped AdS$_{3}$ backgrounds and three-dimensional flat space cosmologies~\cite{Martin:2022duk,Bagchi:2023ilg}. 

Thus far, however, the Selberg zeta function method makes no explicit use of the topological nature of three-dimensional general relativity. One goal of this article is to uncover the topological features in the Selberg zeta function. To do so, we employ a new method for evaluating 1-loop partition functions in three-dimensional gravity. 

\vspace{3mm}

\noindent \textbf{Wilson spool.} The topological nature of three-dimensional Einstein gravity 
makes it suitable to be rewritten as a Chern-Simons theory~\cite{Achucarro:1987vz, Witten:1988hc}. In this context, the Chern-Simons gauge group is equivalent to the spacetime isometry group, and the gauge connections define the spacetime metric. The natural Chern-Simons observables are Wilson loops or lines. 
The  \emph{Wilson spool}---roughly speaking, a collection of Wilson loops winding multiple times along a non-trivial cycle---provides a suitable coupling of matter in the Chern-Simons quantum gravity path integral. In particular, it has been used to compute 1-loop determinants of massive scalar~\cite{Castro:2023dxp, Castro:2023bvo} and higher-spin fields~\cite{Bourne:2024ded} in Euclidean (A)dS$_3$ (for applications to two-dimensional Jackiw-Teitelboim gravity, see~\cite{Fliss:2025sir}). In terms of the background gauge connections $a_L$ and $a_R$, the 1-loop partition function is computed by
\begin{equation}\label{Eq:Wilson_Part}
    Z^{(1)} = \text{exp}\left(\frac{1}{4}\mathbb{W}_{j}\left[a_L,a_R\right]\right)\;,
\end{equation}
where the Wilson spool is defined as~\cite{Castro:2023dxp,Castro:2023bvo,Bourne:2024ded}
\begin{equation}\label{Eq:Wilson}
    \mathbb{W}_{j}\left[a_{L},a_{R}\right] \equiv i\int_{\mathcal{C}}\frac{dz}{z}\frac{\cos{(z/2)}}{\sin{(z/2)}}\text{Tr}_{R_j}\left(\mathcal{P}e^{\frac{z}{2\pi}\oint a_{L}}\right)\text{Tr}_{R_j}\left(\mathcal{P}e^{-\frac{z}{2\pi}\oint a_{R}}\right).
\end{equation}
The specific contour, $\mathcal{C}$, depends on the geometry being considered, and can be derived from a representation-theoretic version of the Denef-Hartnoll-Sachdev method (DHS)~\cite{Denef:2009kn,Denef:2009yy} of evaluating 1-loop determinants using Lorentzian quasinormal modes. The contour integral implements the winding. The traces here are taken over a particular representation $R_j$ associated to the Wilson loops, which can be interpreted as describing a worldline particle associated with the representation $R_{j}$. Crucially, a major advantage of the Wilson spool is that it can be extended off-shell and evaluated within the gravitational path integral, allowing for the computation of quantum gravity corrections in powers of $G_N$ \cite{Castro:2023dxp,Bourne:2024ded}.\footnote{While the construction of the Wilson spool is done on a \emph{fixed} classical background, the relation (\ref{Eq:Wilson_Part}) can be promoted to an off-shell operator by replacing the background gauge connections $a_{L,R}$ to $\mathfrak{su}(2)_{L,R}$ connections $A_{L,R}$ with an associated three-dimensional geometry.}

Importantly, the representations that are suitable for reproducing gravity in de Sitter were shown to be \emph{non-standard} representations~\cite{Castro:2020smu} of the $\mathfrak{su}(2)$ Lie algebra (further developed in~\cite{Castro:2023dxp}). These non-standard representations are infinite-dimensional highest weight representations of $\mathfrak{su}(2)$,
with weights that are not necessarily integer valued, which allow for some of the generators to be non-Hermitian. 
Intriguingly, these representations, suitable for describing de Sitter gravity using Chern-Simons theory, turn out to describe the quasinormal modes on the three-sphere. They differ from the usual representations used to construct quantum fields propagating on a Lorentzian de Sitter spacetime~\cite{Sun:2021thf, Basile:2016aen}. 

Arguably, a disadvantage of the Wilson spool method to compute 1-loop partition functions is that it heavily relies on Chern-Simons machinery. As such, it is difficult to see how such a construction is possible in higher-dimensional gravity, despite the fact the DHS method applies to spacetimes of any dimension. Another goal of this paper is to use the Selberg zeta function as a guide to uncover analogs of the Wilson spool beyond Euclidean (A)dS$_{3}$.  

\vspace{3mm}

\noindent \textbf{Untangling Selberg from the spool.} By a careful interchange of the contour integral and infinite sums over lowest (or highest) weight representations in the Wilson spool (\ref{Eq:Wilson}), we precisely recover the  representation theoretic form of the Selberg zeta function \cite{Bagchi:2023ilg} for arbitrary spin-$s$ bosonic fields on $\mathbb{H}^{3}/\mathbb{Z}$. For example, for the massive scalar on the Euclidean BTZ background, we show 
\beq \frac{1}{4}\mathbb{W}_{j}[a_{L},a_{R}]=-\log(\zeta_{\mathbb{H}^{3}/\mathbb{Z}})\;,\qquad \zeta_{\mathbb{H}^{3}/\mathbb{Z}}=\prod_{\text{descendants}}\langle 1-e^{2\pi\partial_{\phi}}\rangle\;.\eeq
Here $\partial_{\phi}$ is the generator of the quotient group action $\phi\sim \phi+2\pi$ for azimuthal angle $\phi$, with $ e^{2\pi\partial_{\phi}}\equiv \gamma\in\Gamma$ acting on the Hilbert space of fields propagating on $\mathbb{H}^{3}/\Gamma$. The infinite product, meanwhile, is over descendants of a highest weight scalar primary. 

Our findings reveal that,
not only do the Wilson spool and Selberg zeta function compute the same 1-loop determinant, they are structurally equivalent. Indeed, the descendant states are nothing other than the states
associated to the representation $R_{j}$ corresponding to matter propagating along the Wilson loop; this is how the trace over representations associated to the Wilson loops re-organizes into a product over descendants. 
Additionally, the Euler-product over primitive geodesics (\ref{Eq:Selberg3D}) becomes a product over holonomies, and, vice versa, the Wilson spool may be realized as a sum over primitive geodesics.  For higher spin fields in Euclidean BTZ, the spool gives us a starting point for deriving, for the first time, the representation-theory version of the Selberg zeta function in the spinning case. 

\vspace{3mm}

\noindent \textbf{Finding Wilson in the Fredholm determinant.} Since Selberg zeta functions can be constructed for geometries which are not exactly hyperbolic quotients~\cite{Martin:2022duk, Bagchi:2023ilg}, it is natural to ask whether the spool leads to a generalization of the Selberg zeta function for, say, Euclidean dS (or its discrete quotients). Unlike the Euclidean AdS$_{3}$ case, the sum over representations and contour integral in the $S^{3}$ spool cannot be interchanged, and no Euler-product-like zeta function reveals itself. 

Instead, we develop an algorithm for computing the Wilson spool for massive scalar fields on $S^{3}$ by constructing a new trace formula specific to (the resolvent of)  the Fredholm determinant. Historically, the analogous procedure on $\mathbb{H}^{2}/\Gamma$ was how Selberg derived his trace formula and identified the Selberg zeta function. In so doing, we find an appropriate `test function' such that the Wilson spool directly follows from an application of the Poisson resummation formula to the resolvent of the Fredholm determinant. Notably, our procedure straightforwardly extends to higher-dimensional spheres and lens space quotients, for which we propose the analogs of the Wilson spool.

\vspace{3mm} 

\noindent {\bf Outline.} The remainder of this article is as follows. We review the Selberg zeta function and Wilson spool construction in the case of the Euclidean BTZ black hole in Section \ref{SEC:BTZ}. In Section \ref{sec:BTZselbviaWS}, we show the equivalence between the Wilson spool and the representation theoretic version of the Selberg zeta function for arbitrary spin bosons on the Euclidean BTZ background. 
In Section \ref{SEC:DS}, we derive the Wilson spool directly from a trace formula for the resolvent of the Fredholm determinant of a scalar Laplacian on $S^{3}$. We conclude in Section \ref{sec:disc} with a discussion on possible future directions, and here we also present our proposals for the Wilson spool for higher-dimensional spheres and lens space quotients of $S^{3}$. To keep this article self-contained, we include Appendix \ref{app:genzetafuncs} explicitly deriving an important relation between functional and Fredholm determinants, while Appendix \ref{app:traceformbeyondS3} generalizes our trace formula algorithm in Section \ref{SEC:DS} to lens space quotients of $S^{3}$ and odd-dimensional spheres, $S^{d}$. 

\vspace{3mm}

\noindent \textbf{NB.} As our work was in its final stages, we learned of \cite{Bourne:2025azc}, which has some overlap with our results in Section \ref{sec:BTZselbviaWS}. We have coordinated our submissions.

\section{1-loop determinants in AdS$_{3}$: review}\label{SEC:BTZ}

\noindent Here we review the constructions of the Selberg zeta function and Wilson spool in the case of the Euclidean BTZ black hole.

\subsection{Selberg zeta function}\label{sec:BTZSel1}

\noindent We begin by recalling the geometry of the Euclidean BTZ black hole, then will move on to a review of the Selberg zeta function in this case.

\vspace{2mm}

\noindent \textbf{Euclidean BTZ black hole.} The Euclidean BTZ geometry has line element (cf. \cite{Carlip:1994gc, Carlip:1995qv})\footnote{Here units are such that $8G_{\text{N}}=1$. To restore $G_{\text{N}}$, send $M\to 8G_{\text{N}}M$ and $J_{E}\to 4G_{\text{N}}J_{E}$.}
\beq ds_{E}^{2}=N_{E}^{2}dt_{E}^{2}+N_{E}^{-2}dr^{2}+r^{2}(d\phi+N^{\phi}_{E}dt_{E})^{2}\;,\label{eq:EucBTZmetricBL}\eeq
with $\phi\sim\phi+2\pi$, and for lapse $N_{E}$ and shift $N^{\phi}_{E}$
\beq\label{eq:LapseShift} N_{E}=\left(-M+\frac{r^{2}}{\ell_{\text{AdS}}^{2}}-\frac{J_{E}^2}{4r^{2}}\right)^{1/2}\;,\quad N^{\phi}_{E}=\frac{J_{E}}{2r^{2}}\;.\eeq
Here, $\ell_{\text{AdS}}$ is the AdS$_{3}$ length scale, and $M$ and $J_{E}$ are the mass and angular momentum, respectively. The Euclidean lapse $N_{E}$ has roots $r_{\pm}$
\beq r_{\pm}^{2}=\frac{1}{2}M\ell_{\text{AdS}}^{2}\left(1\pm\sqrt{1+\left(\frac{J_{E}}{M\ell_{\text{AdS}}}\right)^{2}}\right)\;,\label{eq:Lorenztianhorradiiv1}\eeq
corresponding to the inner and outer horizon radii, $r_{-}$ and $r_{+}$, respectively, of the (Lorentzian) black hole. Since $r_{+}$ is real and positive, it follows the inner horizon is purely imaginary, $r_{-}=iJ_{E}\ell_{\text{AdS}}/2r_{+}\equiv i|r_{-}|$. 

Famously, the Euclidean BTZ black hole (\ref{eq:EucBTZmetricBL}) can also be realized as a discrete quotient of $\mathbb{H}^3$ \cite{Banados:1992gq},
whose line element in Poincar\'e patch (upper-half plane) coordinates is 
\beq
    ds^2 = \frac{\ell_{\text{AdS}}^2}{z^2}\left(dz^2+dwd\bar{w}\right),
\label{eq:Poincarepatch}\eeq
for $z>0$ and complex coordinate $w\equiv x+iy_{E}$. 
 The metric (\ref{eq:EucBTZmetricBL}) is obtained from the following coordinate transformation (valid only for $r>r_+$):
\beq
\begin{split}
&x= \sqrt{\frac{r^2-r_{+}^{2}}{r^2-r_{-}^{2}}}\cos{\left(\frac{r_{+}t_{E}}{\ell_{\text{AdS}}^{2}}+\frac{\left|r_{-}\right|\phi}{\ell_{\text{AdS}}}\right)}\text{exp}\left(\frac{r_{+}\phi}{\ell_{\text{AdS}}}-\frac{\left|r_{-}\right|t_{E}}{\ell_{\text{AdS}}^2}\right), \\ 
&y_{E}= \sqrt{\frac{r^2-r_{+}^{2}}{r^2-r_{-}^{2}}}\sin{\left(\frac{r_{+}t_{E}}{\ell_{\text{AdS}}^{2}}+\frac{\left|r_{-}\right|\phi}{\ell_{\text{AdS}}}\right)}\text{exp}\left(\frac{r_{+}\phi}{\ell_{\text{AdS}}}-\frac{\left|r_{-}\right|t_{E}}{\ell_{\text{AdS}}^2}\right), \\ 
& z= \sqrt{\frac{r_{+}^{2}-r_{-}^{2}}{r^2-r_{-}^{2}}}\exp\left(\frac{r_{+}\phi}{\ell_{\text{AdS}}}-\frac{\left|r_{-}\right|t_{E}}{\ell_{\text{AdS}}^{2}}\right)\;.
\end{split}
\label{eq:UHPcoordEuc}\eeq

 Regularity of the Euclidean black hole geometry at the horizon $r=r_{+}$, i.e., such that there is no conical singularity, requires the identifications
\beq t_{E}\sim t_{E}+\beta\;,\qquad \phi\sim\phi+\theta\;,\label{eq:Eucbtzids}\eeq
where the (inverse) temperature $\beta$ and angular potential $\theta$\footnote{Note $\theta=i\beta\Omega$ for imaginary (outer) horizon angular velocity $\Omega=-i|r_{-}|/(r_{+}\ell_{\text{AdS}})$ measured at the horizon.} are given by
\beq \beta=\frac{2\pi r_{+}\ell_{\text{AdS}}^{2}}{r_{+}^{2}-r_{-}^{2}}\;,\qquad \theta=\frac{2\pi|r_{-}|\ell_{\text{AdS}}}{r_{+}^{2}-r_{-}^{2}}\;.\eeq
It is straightforward to show the identifications (\ref{eq:Eucbtzids}) applied to \eqref{eq:UHPcoordEuc} yield the identity transformation $(z,w)\to(z,w)$.
Furthermore, the quotient of $\mathbb{H}^3$ is generated by the angular coordinate identification $\phi\sim\phi+2\pi$. Thus, the Euclidean BTZ black hole is understood topologically to be a solid torus $D^2\times S^{1}$ with constant negative curvature, and with \eqref{eq:Eucbtzids} and $\phi\sim\phi+2\pi$ furnishing the contractible and non-contractible cycles, respectively.\footnote{Colloquially, identification \eqref{eq:Eucbtzids} defines a contractible thermal circle, and $\phi\sim\phi+2\pi$ defines a non-contractible spatial circle. In contrast, thermal AdS$_{3}$ (Euclidean AdS$_{3}$ with periodic identification (\ref{eq:Eucbtzids}), where $\beta$ and $\theta$ remain unfixed because there is no black hole horizon) has a non-contractible thermal circle and a contractible spatial circle. Thermal AdS$_{3}$ is thus described by a solid torus, but oppositely filled.} The identification $\phi\sim\phi+2\pi$ gives 
\beq (z,w)\to(|q|^{-1}z,q^{-1}w)\;,\label{eq:idpoincangu}\eeq
where $q\equiv e^{2\pi i\tau_{\text{mod}}}$ with complex modular parameter 
\be\label{eq:taumod}
\tau_{\text{mod}}=\tau_{1}+i\tau_{2}~,
\ee
where 
\be 
\tau_{1}=-|r_{-}|/\ell_{\text{AdS}}~, \indent \tau_{2}=r_{+}/\ell_{\text{AdS}}~.
\ee
The modular parameter $\tau_{\text{mod}}$ is the modulus of the boundary two-torus.

It is also a simple exercise to show the quotient group action associated with the identification $\phi\rightarrow\phi+2\pi$ on coordinates \eqref{eq:UHPcoordEuc} can be written in matrix form (for more details, see \cite{Perry2003SELBERGZF, Bagchi:2023ilg}): 
\begin{equation}\label{eq:GammaOnH3}
    \gamma\begin{pmatrix}x\\y_E\\z\end{pmatrix}=
    \begin{pmatrix}
        e^{2a} & 0 & 0 \\
        0 & e^{2a} & 0 \\
        0 & 0 & e^{2a} \\
    \end{pmatrix}
        \begin{pmatrix}
        \cos 2b & -\sin 2b & 0 \\
        \sin 2b & \cos 2b & 0 \\
        0 & 0 & 1 \\
    \end{pmatrix}
    \begin{pmatrix}x\\y_E\\z\end{pmatrix}.
\end{equation}
 The eigenvalues of the matrix $\gamma$, $e^{2a}$ and $e^{2a\pm 2 i b}$, are the building blocks used by Perry and Williams to construct the Selberg zeta function for the Euclidean BTZ black hole, as we review momentarily.

 The  identifications \eqref{eq:Eucbtzids} can be equivalently generated from a primitive element $\gamma\in \text{SL}(2,\mathbb{C})$, the full isometry group of $\mathbb{H}^{3}$. The M\"{o}bius transformations of $\mathbb{H}^3$ act on coordinates of $\mathbb{H}^3$ in a standard way, for instance see~\cite{Perry2003SELBERGZF}. Specifically, the identification (\ref{eq:idpoincangu}) has\footnote{The identification (\ref{eq:Eucbtzids}) is generated by the identity transformation, i.e., $\gamma$ in (\ref{eq:gammaidtrans}) with $a=0$ and $b=\pi$.} 
\beq \gamma\doteq\begin{pmatrix}e^{a+ib}&0\\0&e^{-(a+ib)}\end{pmatrix}=\begin{pmatrix}q^{-1/2}&0\\0& q^{1/2}\end{pmatrix}\;,\label{eq:gammaidtrans}\eeq
where $a\equiv\pi r_{+}/\ell_{\text{AdS}}=\pi \tau_{2}$ and $b\equiv\pi \left|r_{-}\right|/\ell_{\text{AdS}}=-\pi\tau_{1}$, such that $q=e^{-2(a+ib)}$ and $|q|=(q\bar{q})^{1/2}=e^{-2a}$. 
Consequently, the periodicity of azimuthal angle $\phi$ means the Euclidean BTZ black hole can be understood as a (discrete) quotient of the upper-half plane $\mathbb{H}^{3}$ by subgroup $\Gamma\subset\text{SL}(2,\mathbb{C})$. Presently, $\Gamma$ is generated by $\gamma$, i.e., $\Gamma\equiv\{\gamma^{n}|n\in\mathbb{Z}\}$, such that $\Gamma\simeq\mathbb{Z}$. Thus, the Euclidean BTZ black hole is the orbifold $\mathcal{M}_{\Gamma}\equiv\mathbb{H}^{3}/\Gamma$, for $\Gamma\simeq\mathbb{Z}$: points $p_{1},p_{2}\in\mathbb{H}^{3}$ are identified if $p_{1}=\gamma^{n}\cdot p_{2}$ for $n\in\mathbb{Z}$. Finally, an
element $\gamma\in\Gamma$ is said to be primitive if $\gamma\neq\alpha^{n}$ for any element $\alpha\in\Gamma$ and integer $n>1$.

\vspace{2mm}

\noindent \textbf{Selberg zeta function for BTZ.}  Since the Euclidean BTZ black hole is the hyperbolic quotient, $\mathbb{H}^{3}/\mathbb{Z}$, it has an associated Selberg zeta function, defined as \cite{Perry2003SELBERGZF,Williams:2012ms}
\beq
\zeta_{\mathbb{H}^{3}/\mathbb{Z}}\left(\mathfrak{s}\right)
\equiv\prod_{k_1=0}^{\infty}\prod_{k_2=0}^{\infty}\left[1-e^{-(2a-2ib)k_1}e^{-(2a+2ib)k_2}e^{-2a\mathfrak{s}}\right],
\label{eq:selbzetabtz}\eeq
where $a>0$, $b\in\mathbb{R}$, $k_{1}$ and $k_{2}$ are positive integers, and $\mathfrak{s}\in\mathbb{C}$. The above equation matches the form of \eqref{Eq:Selberg3D}, with $\alpha_1=e^{2ib}$, $\alpha_2=e^{-2ib}$ and $\ell=2a$. The function (\ref{eq:selbzetabtz}) is a zeta function in  that it is an Euler product\footnote{An Euler product expands a Dirichlet series, $\sum_{n=1}^{\infty}a_{n}n^{\mathfrak{s}}$ for $\mathfrak{s}\in\mathbb{C}$, into an infinite product indexed by primes $\mathbb{P}$. For example, the Riemann zeta function has $\zeta(\mathfrak{s})=\prod_{p\in\mathbb{P}}(1-p^{-\mathfrak{s}})^{-1}=\prod_{p\in\mathbb{P}}\left(\sum_{k=0}^{\infty}p^{-k\mathfrak{s}}\right)=\sum_{n=1}^{\infty}n^{-\mathfrak{s}}$. } over `prime geodesics' of (dimensionless) length $\ell(p)=2a=2\pi r_{+}/\ell_{\text{AdS}}$, and, according to Perry and Williams, it converges for all $\mathfrak{s}\in\mathbb{C}$.

A representation theoretic approach for constructing the Selberg zeta function of $\mathbb{H}^3/\Gamma$ was recently developed in \cite{Bagchi:2023ilg}. The starting point is to interpret the primitive group elements $\gamma\in\Gamma$ as acting on the Hilbert space of fields propagating on $\mathbb{H}^3/\Gamma$. To do this, one considers the Killing vectors of $\mathbb{H}^3/\Gamma$ which generate the isometry group $\text{SL}(2,\mathbb{R}) \times \text{SL}(2,\mathbb{R})$. It is well known a particular choice of basis for the Killing vectors results in a set we can label as $\{L_a, \bar{L}_a\}$, with $a=0,\pm1$, which satisfy the algebra
$\left[L_n,L_m\right]=(n-m)L_{n+m}$ (and similarly for generators $\{\bar{L}_{a}\}$) and $[L_{m},\bar{L}_n]=0$. In particular, for the Euclidean BTZ black hole, the generator $\partial_{\phi}$ of the quotient group action $\phi\sim\phi+2\pi$ is 
\begin{equation} \label{eq:partialphi}
2\pi \partial_\phi=2\pi i(\tau_{\text{mod}}L_0-\bar{\tau}_{\text{mod}}\bar{L}_0)~.   
\end{equation}
We can then succinctly express the zeta function \eqref{eq:selbzetabtz} as 
\begin{align}\label{eq:btzrepsel}
    \zeta_{\Gamma} &= \prod_{\text{descendants}}\left<1-\gamma\right>\nonumber \\ 
    &= \prod_{k_1=0}^{\infty}\prod_{k_2=0}^{\infty}\left(1-q^{\frac{\Delta}{2}+k_1}\bar{q}^{\frac{\Delta}{2}+k_2}\right)\;,
\end{align}
where $\gamma=e^{2\pi\partial_{\phi}}$ and the product is over descendants (labeled by $k_1$ and $k_2$) of a highest weight scalar primary of weight $j_L=j_R$.\footnote{Note that $\gamma=e^{2\pi\partial_\phi}$ is a different representation of the same group element written in \eqref{eq:GammaOnH3} and \eqref{eq:gammaidtrans}. Equation \eqref{eq:GammaOnH3} is a three-dimensional representation acting on the upper half-plane, equation \eqref{eq:gammaidtrans} is the two-dimensional fundamental representation, and $\gamma=e^{2\pi\partial_\phi}$ is the infinite-dimensional representation acting on the Hilbert space of states propagating on $\mathbb{H}^3/\mathbb{Z}$ (the highest weight state in this space is a scalar, but the descendants need not be).} These highest weights should be identified with the zeta function parameter 
$\mathfrak{s}$ through $j_L=j_R \equiv 2\mathfrak{s}$.

By now the zeta function (\ref{eq:selbzetabtz}) has long been featured in Euclidean BTZ (or thermal AdS$_{3}$) physics \cite{Perry2003SELBERGZF,Bytsenko:2008wj,Bytsenko:2008ex,Diaz:2008iv,Aros:2009pg,Williams:2012ms,Williams:2014udz,Williams:2015azf,Aros:2016opw,Keeler:2018lza,Keeler:2019wsx,Martin:2019flv}. This was initiated by Perry and Williams, who expressed the BTZ effective action in terms of $\log\zeta_{\mathbb{H}^{3}/\Gamma}$, and found the zeros of the zeta function to coincide with the resonances of scattering operators on the BTZ background \cite{Perry2003SELBERGZF}. Pertinently, the thermal AdS$_{3}$ 1-loop partition function for bosonic and fermionic fields can be cast as a quotient of products of the zeta function \cite{Bytsenko:2008wj,Bytsenko:2008ex,Williams:2014udz,Williams:2015azf}. Specifically, for massive symmetric, traceless, transverse spin-$s$ bosons  (for $s\neq0$),
\beq \text{det}\left(-\nabla_{(s)}^{2}+m_{s}^{2}\ell_{\text{AdS}}^{2}\right)=\zeta_{\mathbb{H}^{3}/\Gamma}\left(\Delta_{s}+\frac{isb}{a}\right)^2\zeta_{\mathbb{H}^{3}/\Gamma}\left(\Delta_{s}-\frac{isb}{a}\right)^2\;.\label{eq:functionaldetAdS}\eeq
Here $\nabla^{2}_{(s)}$ is the Laplace-Beltrami operator acting on symmetric, transverse, traceless (STT) tensor fields, and $\Delta_{s}$ is the conformal dimension of the CFT$_{2}$ primary dual to the spin-$s$ field on AdS$_{3}$,
\beq \Delta_{s}\equiv 1+\sqrt{(s-1)^{2}+m^{2}\ell_{\text{AdS}}^{2}}~,\label{eq:confDmass}\eeq
where $m$ is the standard mass that related to the effective mass $m_{s}$ via \cite{Datta:2011za}
\be 
m_s^2 \ell_{\text{AdS}}^2 = m^2 \ell_{\text{AdS}}^2 + s(s-3)~.
\label{eq:effmass}\ee
It follows that the 1-loop partition function is \cite{Williams:2015azf}\footnote{To arrive to the second line it is useful to know $q^{j_{L}+k_{1}}\bar{q}^{j_{R}+k_{2}}=e^{i(b+ia)(\Delta_{s}+s+2k_{1})}e^{-i(b-ia)(\Delta_{s}-s+2k_{1})}$.}
\begin{align}
Z(\Delta_{s})&=\;\biggr|\zeta_{\mathbb{H}^{3}/\mathbb{Z}}\left(\Delta_{s}+\frac{isb}{a}\right)\biggr|^{-2}\nonumber \\
&=\prod_{k_{1},k_{2}=0}^{\infty}\left(1-q^{j_{L}+k_{1}}\bar{q}^{j_{R}+k_{2}}\right)^{-1}\left(1-q^{j_{R}+k_{1}}\bar{q}^{j_{L}+k_{2}}\right)^{-1}\;,
\end{align}
where in the first line  $\zeta_{\mathbb{H}^{3}/\mathbb{Z}}(\bar{\mathfrak{s}})=\overline{\zeta_{\mathbb{H}^{3}/\mathbb{Z}}(\mathfrak{s})}$ was used, and 
\beq j_{L}\equiv \frac{(\Delta_{s}+s)}{2}\;,\quad j_{R}\equiv \frac{(\Delta_{s}-s)}{2}\;.\label{eq:jljrAdS}\eeq
 Note that for scalar fields ($s=0$), where $j_{L}=j_{R}=\Delta/2$, the functional determinant (\ref{eq:functionaldetAdS}) reduces to a single zeta function.

\subsection{Wilson spool}\label{sec:wsads}

\noindent The Wilson spool relies on the Chern--Simons formulation of gravity \cite{Achucarro:1987vz,Witten:1988hc} (see, e.g.,  Appendix A of \cite{Bourne:2024ded} for a review). 
In this formalism, the topological nature of gravity in three dimensions allows the metric $g_{\mu\nu}$ (expressed in terms of vielbeins $e_\mu^a$ and spin connections $\omega_\mu^{ab}$) to be entirely recast in terms of gauge connections, $a_L$ and $a_R$. Specifically, the background gauge connections associated with the Euclidean BTZ black hole (\ref{eq:EucBTZmetricBL}) are~\cite{Gutperle:2011kf}\footnote{Note that connections (\ref{eq:BTZholonomies}) are the Euclideanized background gauge connections that build the \emph{Lorentzian} BTZ geometry. Since we focus on applications to $1$-loop partition functions, we will always work in Euclidean signature. For simplicity we choose to work with the $SO(2,2)$ Chern-Simons reformulation of Lorentzian AdS$_{3}$ gravity analytically continued to Euclidean, as opposed to starting from the $SL(2,\mathbb{C})$ Chern-Simons formulation of Euclidean AdS$_{3}$ gravity, to avoid subtleties that arise with a complex gauge group.}
\begin{align}\label{eq:BTZholonomies}
    a_L &= L_0dr +\left(e^{r}L_+-\frac{1}{4}\frac{\left(r_++i|r_-|\right)^2}{\ell_{\text{AdS}}^2}e^{-r}L_-\right)\left(-idt_E+d\phi\right)~, \nonumber \\
    a_R &= -\bar{L}_0dr +\left(e^{r}\bar{L}_--\frac{1}{4}\frac{\left(r_+-i|r_-|\right)^2}{\ell_{\text{AdS}}^2}e^{-r}\bar{L}_{+}\right)\left(idt_E+d\phi\right)~.
\end{align}
Here, $L_0,L_\pm$ (and barred versions) are generators of $\mathfrak{sl}(2,\mathbb{R})$. The full gauge group $SO(2,2) \simeq\text{SL}(2,\mathbb{R})_L \times \text{SL}(2,\mathbb{R})_R$ is nothing other than the isometry group of the (Lorentzian) spacetime.

A natural observable in Chern-Simons theory is a \emph{Wilson loop}. In terms of the gauge connections, a Wilson loop is defined as 
\be\label{eq:WilsonLoop}
W_{\mathsf{R_{\Delta, s}}} \equiv  \mbox{Tr}_\mathsf{R_{\Delta, s}} \left(\mathcal{P} e^{\oint_\gamma a_L}\mathcal{P} e^{\oint_\gamma a_R}\right)~.
\ee 
The Wilson loop explicitly depends on a representation of the full gauge algebra, $\mathfrak{so}(2,2)$. As in Section~\ref{sec:BTZSel1}, we take this representation to consist of a pair of highest- or lowest-weight representations of $\mathfrak{sl}(2,\mathbb{R})$,
which we denote $\mathsf{R}_{\Delta, s}$ since both the conformal dimension and spin, ($\Delta$, $s$),\footnote{Note that $\Delta$ also depends on $s$ as in~\eqref{eq:confDmass}, but we will henceforth drop the subscript.} uniquely specify such a representation. 
Equivalently, through \eqref{eq:jljrAdS}, the representation can be specified by the pair of highest (lowest) weights $(j_L,j_R)$ of each copy of $\mathfrak{sl}(2,\mathbb{R})$.
Physically, representation for the Wilson loop describes the type of probe matter present on a fixed gravitational background set up by the connections, $a_L$ and $a_R$. Given the choice $\mathsf{R}_{\Delta, s}$, the Wilson loop describes a massive, spinning field with mass $m$ and spin $s$ set by the Casimirs, $c_2=j(j-1)$ and highest weights of the two $\mathfrak{sl}(2,\mathbb{R})$ representations.

Inside the trace of the Wilson loop~\eqref{eq:WilsonLoop} is a path ordered exponential of the gauge connection along a closed curve $\gamma$ in the bulk spacetime. Specifically, 
\begin{equation}\label{eq:holBTZ}
    \mathcal{P}e^{\oint_{\gamma}a_L} = g_L^{-1}e^{2\pi i h_L L_3}g_{L}~, \indent  \mathcal{P}e^{\oint_{\gamma}a_R} = g_R^{-1}e^{2\pi i h_R \bar{L}_3}g_{R}~,
\end{equation}
where $g_{L/R}$ are fixed group elements that drop out under the trace, and where the holonomies\footnote{Geometrically, the holonomy group measures the failure of a single coordinate patch to entirely extend around a non-contractible cycle in spacetime and are quantified by the Wilson loops of the gauge connection. The path ordered exponentials are specified by a group element that can be identified by constants $h_{L,R}$. Hence, we will refer to both as holonomies.}
for the Euclidean BTZ black hole are 
\begin{equation}\label{eq:btzhol}
    h_L = \tau_{\rm mod}~, \quad h_R =\bar{\tau}_{\rm mod}~,
\end{equation}
for modular parameter $\tau_{\text{mod}}$ defined in~\eqref{eq:taumod}.

A Wilson loop on the Euclidean BTZ black hole background was shown to have an interesting gravitational interpretation when it wraps a nontrivial object in the bulk. For instance, a Wilson loop wrapping the horizon of a black hole was generically shown to compute the black hole's thermal entropy~\cite{Ammon:2013hba}.
Intuitively, a \emph{Wilson spool} is constructed from a collection of Wilson loops winding around a nontrivial cycle~\cite{Castro:2023dxp,Castro:2023bvo,Bourne:2024ded,Fliss:2025sir}. On a fixed background geometry, the spool computes 1-loop determinants corresponding to the probe matter. The definition of the spool can be extended off-shell, through evaluation of the path integral, thus providing a way to couple quantum gravity to matter in Chern-Simons gravity.

The construction of the spool is inspired by a representation-theoretic version of the Denef-Hartnoll-Sachdev (DHS) method~\cite{Denef:2009kn,Denef:2009yy} for computing 1-loop determinants. For illustrative purposes, consider the 1-loop determinant $Z_{\Delta, s}$ for a real massive spin-$s$ field,
\begin{equation}\label{eq:detspinnew}
    Z_{\Delta, s} = \text{det}\left(-\nabla_{(s)}^2+m_{s}^2 \ell_{\text{AdS}}^2\right)^{-1/2},
\end{equation}
where $\nabla_{(s)}^{2}$ is the Laplace-Beltrami operator and $m_s$ is the effective mass, with an associated conformal dimension (\ref{eq:confDmass}). In the usual DHS method, one uses Weierstrass factorization to express meromorphic functions, such as $Z_{\Delta, s}$, as a product over its zeros and poles (up to an overall entire function). By~\eqref{eq:detspinnew}, poles occur when
\be -\nabla_{(s)}^2+m_{s}^2\ell_{\text{AdS}}^2=0~,\label{eq:eom}\ee
which precisely correspond to tuning $\Delta_{(s)}$ such that the Lorentzian quasinormal modes Wick rotate to regular Euclidean modes, where the QNM and thermal Matsubara frequencies coincide \cite{Castro:2017mfj} (see also \cite{Keeler:2018lza,Keeler:2019wsx,Grewal:2022hlo}).

It is worth reviewing the steps leading to the Wilson spool using the representation theory DHS construction of $Z_{\Delta, s}$, since they will be useful to recast the spool in terms of the representation-theoretic form of the Selberg zeta function in Section~\ref{sec:BTZselbviaWS}. First, a state in the Chern-Simons representation $\mathsf{R}_{\Delta_s}$ is considered to contribute a pole to the 1-loop determinant if it obeys three key conditions~\cite{Bourne:2024ded,Fliss:2025sir}: \\

\noindent \textbf{``Mass shell'' condition:} Let $\ket{\psi}\in\mathsf{R}_{\Delta, s}$ be a state in a representation $\mathsf{R}_{\Delta, s}$.  Written in terms of the Casimir operator,\footnote{
Recall that by expressing the Casimir operator in terms of the $\mathfrak{sl}(2,\mathbb{R})_L\oplus\mathfrak{sl}(2,\mathbb{R})_R$ Killing vectors $\zeta_a\,\bar{\zeta}_a$, we have
$$
    2c_{2,L}+ 2c_{2,R} = \nabla^{2}_{(s)}+s(s+1).
$$
} (\ref{eq:eom}) becomes 
\begin{equation}\label{eq:caseqspin}
    \left(c_{2,L} + c_{2,R}\right)\ket{\psi} = \frac{1}{2}\left(\Delta \left(\Delta -2\right) + s^2\right)\ket{\psi}.
\end{equation}
Let $\mathcal{R}_{\Delta, s}$ denote the set of all representations satisfying (\ref{eq:caseqspin}). Since this Casimir condition remains true for both highest and lowest weight representations, we have 
\begin{equation}\label{eq:reps}
    \mathcal{R}_{\Delta, s} = \mathcal{R}_{\Delta, s}^{\text{HW}}\cup \mathcal{R}_{\Delta, s}^{\text{LW}}~,
\end{equation} 
where $\mathcal{R}^{\text{HW}}_{\Delta, s}, \mathcal{R}^{\text{LW}}_{\Delta, s}$ are the set of highest and lowest weight representations of $\mathfrak{sl}(2,\mathbb{R})_L\oplus\mathfrak{sl}(2,\mathbb{R})_R$. For fixed $j_L,j_R$, there are two possible representations that contribute to each:
\begin{align}\label{eq:reps2}
\mathcal{R}^{\text{HW}}_{\Delta, s} &= \{\mathsf{R}^{\text{HW}}_{j_L}\otimes\mathsf{R}^{\text{HW}}_{j_R},\mathsf{R}^{\text{HW}}_{j_R}\otimes\mathsf{R}^{\text{HW}}_{j_L}\}~,\nonumber\\
\mathcal{R}^{\text{LW}}_{\Delta, s}&=\{\mathsf{R}^{\text{LW}}_{j_L}\otimes\mathsf{R}^{\text{LW}}_{j_R},\mathsf{R}^{\text{LW}}_{j_R}\otimes\mathsf{R}^{\text{LW}}_{j_L}\}~.
\end{align}
All of these different representations contribute poles to $Z_{\Delta, s}$. Finally, note that the shadow map, $\Delta \rightarrow \bar{\Delta} = 2 - \Delta$, also satisfies the Casimir equation (\ref{eq:eom}). When Dirichlet boundary conditions are imposed on the solutions contributing to the 1-loop partition function \eqref{eq:detspinnew}, only $\Delta$ needs to be considered.\footnote{Note that the shadow representations are necessary to include in the case of de Sitter in Section~\ref{SEC:DS}.}

\vspace{3mm} 

\par
\noindent \textbf{Parallel transport invariance:}
The first condition requires that a field $\Phi$ in a representation $\mathsf{R}_L\otimes\mathsf{R}_R\in\mathcal{R}_{\Delta_s}$ remain invariant under parallel transport round a closed cycle  $\gamma$: 
\begin{equation}\label{eq:pards}
    \Phi_{f} = \mathsf{R}_L
    \left(\mathcal{P}e^{\oint_{\gamma}a_L}\right) \Phi_i \mathsf{R}_R
    \left(\mathcal{P}e^{-\oint_{\gamma}a_R}\right).
\end{equation}
For the BTZ black hole, this is not trivially satisfied around the non-contractible $\phi$-cycle, $\gamma_\phi$. This gives rise to the constraint
\begin{equation}\label{eq:holcon}
    \lambda_Lh_L-\lambda_Rh_R = |n|~,\quad n\in\mathbb{Z}~,
\end{equation}
for holonomies (\ref{eq:btzhol}) and where $(\lambda_L,\lambda_R)$ is a particular weight of the representation. 

\vspace{3mm}

\par
\noindent \textbf{Globally regular solutions:}
The second condition is the requirement of globally regular solutions, phrased in terms of representation theory: for a state in a representation to contribute to the poles of the 1-loop partition function \eqref{eq:detspinnew}, the representation  of the algebra must lift to a representation of the group. For the BTZ black hole, since every representation of $\mathfrak{sl}(2, \mathbb{R})$ lifts to a representation of $\text{SL}(2, \mathbb{R})$, this condition will be trivially satisfied for all $\mathsf{R}_{L} \otimes \mathsf{R}_{R} \in \mathcal{R}_{\Delta, s}$.
\\

With these conditions, we may express the 1-loop partition function $Z_{\Delta, s}$ in the DHS-inspired product form as
\begin{equation}\label{eq:DHSspin}
    Z_{\Delta, s} = \prod_{\mathcal{R}_{\Delta_s}}\prod_{\left(\lambda_L,\lambda_R\right)}\prod_{n\in\mathbb{Z}}\left(|n|-\lambda_{L}h_{L}+\lambda_{R}h_{R}\right)^{-1/4}\left(|n|+\lambda_{L}h_{L}-\lambda_{R}h_{R}\right)^{-1/4}~.
\end{equation}
To recast this product into the contour integral form that defines the Wilson spool $\mathbb{W}_{j_L,j_R}$, take the logarithm of $Z_{\Delta, s}$, convert the products to summations, and subsequently rewrite the logarithms as integrals using the Schwinger trick,
\begin{equation}\label{eq:Schwinger}
    \log{M} = -\int_{\cross}^{\infty}\frac{dz}{z}e^{-zM}~.
\end{equation}
Applying these steps and evaluating the sum over $n\in\mathbb{Z}$, the 1-loop determinant is
\begin{align}\label{eq:btzw1}
    \log{Z_{\Delta, s}} &= \frac{1}{4}\sum_{\mathcal{R}_{\Delta_s}}\int_{\cross}^{\infty}\frac{dz}{z}\frac{\cosh{(z/2)}}{\sinh{(z/2)}}\sum_{\left(\lambda_{L},\lambda_{R}\right)}e^{z(\lambda_{L}h_L-\lambda_{R}h_{R})} \nonumber \\ 
    &+\frac{1}{4}\sum_{\mathcal{R}_{\Delta_s}}\int_{\cross}^{\infty}\frac{dz}{z}\frac{\cosh{(z/2)}}{\sinh{(z/2)}}\sum_{\left(\lambda_{L},\lambda_{R}\right)}e^{-z(\lambda_{L}h_L-\lambda_{R}h_{R})}~.
\end{align}
The symmetry of the integrand can be exploited by flipping the bounds of integration for the second line, and taking $z\rightarrow-z$ to express the 1-loop determinant as a single integral over the real line. Additionally, we must apply an $i\epsilon$ prescription to both regulate the divergence at $z=0$, and to ensure the series over the weight $(\lambda_L,\lambda_R)$ converges. Whether to take $z\rightarrow z+i\epsilon$ or $z\rightarrow z-i\epsilon$ will depend on the representation, since it is the weight that determines the overall sign in the exponential. The convergence of the sum in (\ref{eq:btzw1}) is determined by the imaginary parts of the holonomies, due to the $i\epsilon$ deformation. Notably,
$\operatorname{Im}(h_L) > 0$, and $\operatorname{Im}(h_R) < 0.$
The relative minus sign in front of $h_R$ causes both $k_1 h_L$ and $-k_2 h_R$ terms to contribute negative real parts to the exponent, leading to exponential decay and convergence.

\begin{figure}[t!]
    \centering
    \resizebox{1.0\linewidth}{!}{
    \begin{tikzpicture}[scale=1.1, decoration={markings,
        mark=at position 0.5cm with {\arrow[line width=2.5pt]{>}},
        mark=at position 2cm with {\arrow[line width=2.5pt]{>}},
        mark=at position 3.5cm with {\arrow[line width=2.5pt]{>}},
        mark=at position 5.5cm with {\arrow[line width=2.5pt]{>}},
        mark=at position 7.85cm with {\arrow[line width=2.5pt]{>}},
        mark=at position 9cm with {\arrow[line width=2.5pt]{>}},
        mark=at position 11cm with {\arrow[line width=2.5pt]{>}}
    }]
        \begin{scope}[xshift=0cm]
            \draw[help lines,<->,line width=2pt] (-6.5,0) -- (6.5,0);
            \draw[help lines,<->,line width=2pt] (0,-4.5) -- (0,4.5);
            \path[draw,line width=2pt,postaction=decorate, black] (-.75,-4.5) -- (-.75,4.5);
            \path[draw,line width=2pt,postaction=decorate, black] (.75,-4.5) -- (.75,4.5);
            \node[below right] at (4.3,-.3) {\Huge{ Re$(z)$}};
            \node[above left] at (1,4.5) {\Huge{Im$(z)$}};
            \node at (-2,3) {\Huge $\mathcal{C}_-$};
            \node at (2,3) {\Huge$\mathcal{C}_+$};
            \foreach \x in {-6,-4.5,-3,-1.5,0,1.5,3,4.5,6} {
                \filldraw[violet] (\x,0) circle (2pt);
            }
            \node at (0,-5.2) {\Huge{(a)}};
        \end{scope}

        \begin{scope}[xshift=13.5cm]
            \draw[help lines,<->,line width=2pt] (-6.5,0) -- (6.5,0);
            \draw[help lines,<->,line width=2pt] (0,-4.5) -- (0,4.5);
            \path[draw,line width=2pt,postaction=decorate, black] (0.37,-4.5) -- (0.37,4.5);
            \path[draw,line width=2pt,postaction=decorate, black] (1.2,-4.5) -- (1.2,4.5);
            \node[below right] at (4.3,-.3) {\Huge Re$(z)$};
            \node[above left] at (1,4.5) {\Huge Im$(z)$};
            \node at (2,3) {\Huge$\mathcal{C}$};
            \foreach \x in {-6,-4.5,-3,-1.5,0,1.5,3,4.5,6} {
                \filldraw[violet] (\x,0) circle (2pt);
            }
            \node at (0,-5.2) {\Huge{(b)}};
        \end{scope}

        \begin{scope}[xshift=27cm]
            \draw[help lines,<->, line width=2pt] (-6.5,0) -- (6.5,0);
            \draw[help lines,<->,line width=2pt] (0,-4.5) -- (0,4.5);
            \path[draw,line width=1.5pt,postaction=decorate, black] (6.5,-1) -- (2,-1) arc (270:90:1) -- (6.5,1);
            \path[draw,line width=1.5pt,postaction=decorate, black] (6.5,-0.5) -- (1.91,-0.5) arc (270:90:0.5) -- (6.5,0.5);
            \node[below right] at (6.75,.5) {\Huge Re$(z)$};
            \node[above left] at (1,4.5) {\Huge{Im$(z)$}};
            \node at (3,2) {\Huge{$\mathcal{C}$}};
            \foreach \x in {-6,-4.5,-3,-1.5,0,1.85,3,4.5,6} {
                \filldraw[violet] (\x,0) circle (2pt);
            }
            \node at (0,-5.2) {\Huge{(c)}};
        \end{scope}
    \end{tikzpicture}
    }
    \caption{(a) The contour when split into lowest weight representations, $\mathcal{C}_+$, and highest weight representations, $\mathcal{C}_-$. (b) The  contour expressed in terms of only highest weight representations, $\mathcal{C}=2\mathcal{C}_+$. (c) The contour $\mathcal{C}$ deformed to enclose poles that lie on the real axis. }
    \label{fig:Contour}
\end{figure}
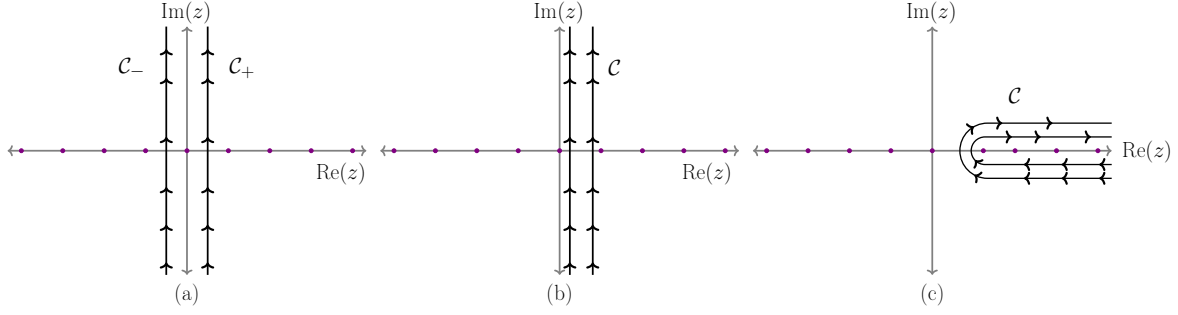

 Using the decomposition \eqref{eq:reps}, the integrals split into two contours running above and below the real axis, corresponding to the contributions from highest- and lowest-weight representations. Taking $z\rightarrow-iz$, the partition function (\ref{eq:btzw1}) becomes
\begin{align}\label{eq:1loopWSde}
    \log{Z_{\Delta, s}} &= \frac{i}{4}\sum_{\mathcal{R}_{\Delta_s}^{\text{LW}}}\int_{\mathcal{C}_+}\frac{dz}{z}\frac{\cos{(z/2)}}{\sin{(z/2)}}\sum_{\left(\lambda_L,\lambda_R\right)}e^{iz(\lambda_{L}h_{L}-\lambda_{R}h_{R})}\nonumber \\ 
    &+ \frac{i}{4}\sum_{\mathcal{R}_{\Delta_s}^{\text{HW}}}\int_{\mathcal{C}_-}\frac{dz}{z}\frac{\cos{(z/2)}}{\sin{(z/2)}}\sum_{\left(\lambda_L,\lambda_R\right)}e^{iz(\lambda_{L}h_{L}-\lambda_{R}h_{R})}~,
\end{align}
where $\mathcal{C}_{+}$ and $\mathcal{C}_{-}$ run to the right and left of the imaginary axis, as shown in Figure \ref{fig:Contour}.

We can collapse the two contour integrals in (\ref{eq:1loopWSde}) into a single integral. To wit, consider  the $\mathcal{C}_-$ integral over a highest weight representation. The only difference between a weight $(\lambda_L, \lambda_R)$ in a lowest or highest weight representation is the overall sign of the weight. As a result, evaluating the highest weight integral over $\mathcal{C}_-$ is equivalent to evaluating the integral over lowest weight representations on contour $\mathcal{C}_+$. We  adopt the convention of expressing the 1-loop determinant (\ref{eq:1loopWSde}) in terms of only lowest weight representations. Consequently, 
 \be
 \log{Z_{\Delta, s}} = \frac{1}{4} \mathbb{W}_{j_L,j_R}~,
 \ee
 where the Wilson spool is defined as
\begin{equation}\label{eq:spooladsZ1}
    \mathbb{W}_{j_L,j_R} \equiv i\sum_{\mathcal{R}_{\Delta_s}^{\text{LW}}}\int_{\mathcal{C}}\frac{dz}{z}\frac{\cos{(z/2)}}{\sin{(z/2)}}\text{Tr}_{\mathsf{R}_L}\left(\mathcal{P}e^{\frac{z}{2\pi}\oint_{\gamma_\phi} a_L}\right)\text{Tr}_{\mathsf{R}_R}\left(\mathcal{P}e^{-\frac{z}{2\pi}\oint_{\gamma_\phi} a_R}\right).
\end{equation}
Now, $\mathcal{C}=2 \mathcal{C}_{+}$, as shown in Figure \ref{fig:Contour}. The poles of the integrand are located on the real axis at $z=2\pi n$ for $n\in\mathbb{Z}$, and the integral can be evaluated as a sum over residues of these poles. 
\par

\section{Selberg zeta function = Wilson spool for BTZ}\label{sec:BTZselbviaWS} 

\noindent Here, starting from the Wilson spool, we derive the representation theoretic expression for the Selberg zeta function for massive symmetric, transverse, traceless spin-$s$ fields on the (Euclidean) rotating BTZ black hole background. 

\subsection{Scalar case} \label{sec:scalarselb}

\noindent  We begin with a massive scalar field ($s=0$), 
where $j_L=j_R \equiv j$.
The 1-loop partition function computed using the spool (\ref{eq:spooladsZ1}) simplifies to
\begin{equation}\label{eq:spoolads}
    \log{Z_\Delta} = \frac{i}{4}\int_{\mathcal{C}}\frac{dz}{z}\frac{\cos{(z/2)}}{\sin{(z/2)}}\text{Tr}_{\mathsf{R}_j}\left(\mathcal{P}e^{\frac{z}{2\pi}\oint_{\gamma_\phi} a_L}\right)\text{Tr}_{\mathsf{R}_j}\left(\mathcal{P}e^{-\frac{z}{2\pi}\oint_{\gamma_\phi} a_R}\right),
\end{equation}
where $\mathsf{R}_j$ is a lowest weight representation of $\mathfrak{sl}(2,\mathbb{R})$ with the lowest weight labeled by $j$. As reviewed above,  a weight $(\lambda_L,\lambda_R) \in   \mathsf{R}^{\text{LW}}_{j}\otimes\mathsf{R}^{\text{LW}}_{j},$
can be expressed as ascendants of the lowest weight state labeled by $j$. Using the decomposition (\ref{eq:1loopWSde}) gives 
\begin{align}\label{eq:spool0}
    \log{Z_\Delta} 
    &= \frac{i}{4}\int_{\mathcal{C}}\frac{dz}{z}\frac{\cos{(z/2)}}{\sin{(z/2)}}\sum_{k_1=0}^{\infty}\sum_{k_2=0}^{\infty}e^{iz(h_L(j+k_1)-h_R(j+k_2))}~,
\end{align}
where we implemented the sum over lowest weights.

The derivation of the representation theory form of the Selberg zeta function essentially follows from an interchange 
of the infinite sums and contour integral in (\ref{eq:spool0}). It is worth performing this exchange carefully as it is not always possible (in fact, as we show in Section \ref{SEC:DS}, this interchange is not allowed for the Euclidean dS$_{3}$ 1-loop partition function).
To this end, following a standard prescription in complex analysis (cf. \cite{brown2009complex}), we decompose the infinite sums over $k_1$ and $k_2$ into finite sums along with a remainder
$\rho_{M,N}(z)$. Specifically, 
\beq
\begin{split}
 \sum_{k_{1},k_{2}=0}^{\infty}e^{iz(h_L(j+k_1)-h_R(j+k_2))}&
 =\sum_{k_1=0}^{M-1}\sum_{k_2=0}^{N-1}e^{iz(h_L(j+k_1)-h_R(j+k_2))}+\rho_{M,N}(z)\;,
 \end{split}
 \label{eq:remain}\eeq
with 
\beq \label{eq:remainder}
\begin{split} 
\rho_{M,N}(z)&\equiv e^{izj(h_{L}-h_{R})}\biggr[\sum_{k_{1}=0}^{M-1}e^{izk_{1}h_{L}}\sum_{k_{2}=N}^{\infty}e^{-izk_{2}h_{R}}+\sum_{k_{2}=0}^{N-1}e^{-izk_{2}h_{R}}\sum_{k_{1}=M}^{\infty}e^{ik_{1}h_{L}}\\
&+\sum_{k_{1}=M}^{\infty}e^{izk_{1}h_{L}}\sum_{k_{2}=N}^{\infty}e^{-izk_{2}h_{R}}\biggr]\;,
\end{split}
\eeq
for positive integers $M,N$. Evaluating the sums in the remainder gives\footnote{Recall $\sum_{k=M}^{\infty}e^{kx}=\frac{e^{Mx}}{1-e^{x}}$ and $\sum_{k=0}^{M-1}e^{kx}=\frac{e^{Mx}-1}{e^{x}-1}$.}
\beq 
\begin{split} 
\rho_{M,N}(z)&=
\frac{e^{\frac{iz}{2}[(2j-1)(h_{L}-h_{R})+2(Mh_{L}-Nh_{R})]}}{4\sin\left(\frac{zh_{L}}{2}\right)\sin\left(\frac{zh_{R}}{2}\right)}\left(e^{-izh_{L}M}+e^{izh_{R}N}-1\right)\;.
\end{split}
\label{eq:remainBTZsca}\eeq
With the decomposition (\ref{eq:remain}), the finite sums may be interchanged with the integral in (\ref{eq:spool0}). Subsequently taking the limit $M,N \rightarrow \infty$ gives,
\begin{align}\label{eq:spool4}
    \int_{\mathcal{C}}\frac{dz}{z}\frac{\cos{(z/2)}}{\sin{(z/2)}}\sum_{k_1,k_2=0}^{\infty}e^{iz(h_L(j+k_1)-h_R(j+k_2))} &= \sum_{k_1,k_2=0}^{\infty}\int_{\mathcal{C}}\frac{dz}{z}\frac{\cos{(z/2)}}{\sin{(z/2)}}e^{iz(h_L(j+k_1)-h_R(j+k_2))}\nonumber \\  &+\lim_{M,N\rightarrow\infty}\int_{\mathcal{C}}\frac{dz}{z}\frac{\cos{(z/2)}}{\sin{(z/2)}}\rho_{M,N}(z)~.
\end{align}
Thus, interchanging the sum and integral is valid provided the remainder (\ref{eq:remainBTZsca}) vanishes in the limit $M,N\rightarrow \infty$. 

For Euclidean BTZ, the holonomies (\ref{eq:btzhol}) have $ \operatorname{Im}(h_L) \neq 0 $ and $ \operatorname{Im}(h_R) \neq 0 $, such that the remainder will not contribute any poles that lie inside $\mathcal{C}$. The only poles lying inside $\mathcal{C}$ are first order poles located at $z=2\pi n$ for $n\in\mathbb{Z}$. Therefore, we can evaluate the contour integral for the remainder term by summing over the residues of the poles enclosed by $\mathcal{C}$. Convergence depends only on the convergence of the exponential terms in (\ref{eq:spool4}). The overall exponential will be damped, and in the limit $M,N\rightarrow \infty$, we obtain
\begin{align}\label{eq:spool6}
    \lim_{M,N\to\infty}\int_{\mathcal{C}}\frac{dz}{z}\frac{\cos{(z/2)}}{\sin{(z/2)}}\rho_{M,N}(z)dz = -4i\lim_{M,N\rightarrow\infty}\sum_{n=1}^{\infty}\frac{\rho_{M,N}(2\pi n)}{n} = 0~.
\end{align}
Thence,
the infinite sums and the integral in (\ref{eq:spool0}) can be interchanged, and the 1-loop determinant 
takes the form
\begin{align}
    \log{Z_\Delta} &= \frac{i}{4}\sum_{(\lambda_L,\lambda_R)}\int_{\mathcal{C}}\frac{dz}{z}\frac{\cos{(z/2)}}{\sin{(z/2)}}e^{iz(\lambda_L h_L-\lambda_R h_R)}\nonumber \\ 
    &= \log{\prod_{(\lambda_L,\lambda_R)}\left[1-e^{2\pi i\left(\lambda_Lh_L-\lambda_R h_R\right)}\right]^{-1}}.
\end{align}

It is now trivial to complete our derivation of the representation-theoretic form of the Selberg zeta function (\ref{eq:btzrepsel}). Using the relation (\ref{eq:selr}) between the Selberg zeta function and the 1-loop determinant and identifying the appropriate generator through \eqref{eq:partialphi}, we find
\begin{align}\label{eq:RepTheoryDerive}
    \zeta_{\Gamma} (\Delta) &= \prod_{(\lambda_L,\lambda_R)}\left(1-e^{2\pi i(h_L\lambda_L-\lambda_R h_R)}\right)\nonumber \\ 
    &= \prod_{\text{descendants}}\left<1-e^{2\pi\partial_{\phi}}\right>~,
\end{align}
precisely recovering the representation theory  construction of the BTZ Selberg zeta function established in \cite{Bagchi:2023ilg}.

In summary, we have seen that the logarithm of the Selberg zeta function can be directly obtained from the Wilson spool, through a simple interchange of a sum and integral. In equating the two, there are two key relations we would like to highlight:\\

\noindent {\bf States are those associated to the Wilson spool}: First, the states at play in the representation theory expression for the Selberg zeta function are nothing other than the states associated to the representation $\mathsf{R}_\Delta$, corresponding to matter propagating along the Wilson loop. The states in $\mathsf{R}_\Delta$ are summed over in the Wilson spool through the trace. Since the Wilson spool is logarithm of the Selberg zeta function, this corresponds to the product over descendants in the representation theory form of the zeta function. \\ 

\noindent{\bf Geodesic lengths are swapped for holonomies}: As was reviewed in Section~\ref{SEC:BTZ}, the Selberg zeta function is a zeta function constructed from the lengths of primitive geodesics, rather than prime numbers. In the language of the Wilson spool, the holonomies generalize geodesic lengths. To see this, recall  \eqref{eq:btzhol} expressed in terms of the (dimensionless) geodesic length, $\ell(p)=2\pi r_{+}/\ell_{\text{AdS}}$,
\begin{equation}\label{hol}
    h_{L} = -\frac{|r_{-}|}{\ell_{\text{AdS}}}+\frac{i\ell(p)}{2\pi}~,\indent h_{R} = -\frac{|r_{-}|}{\ell_{\text{AdS}}}-\frac{i\ell(p)}{2\pi}~.
\end{equation}
The zeta function~\eqref{eq:RepTheoryDerive} for BTZ  that was directly obtained from the Wilson spool, 
\begin{equation}\label{eq:selrep}
    \zeta_{\Gamma}(\Delta) =\prod_{k_1=0}^{\infty}\prod_{k_2=0}^{\infty}\left(1-e^{2\pi i(h_{L}(j+k_1)-h_R(j+k_2))}\right)~,
\end{equation}
 thus reduces to the geometric form of the Selberg zeta function (\ref{eq:selbzetabtz}) expressed in terms of the primitive geodesic length $\ell(p)$, 
\begin{equation}
    \zeta_{\Gamma}(\Delta) = \prod_{k_1=0}^{\infty}\prod_{k_2=0}^{\infty}\left(1-e^{-\left(2\pi i\frac{|r_{-}|}{\ell_{\text{AdS}}}+\ell(p)\right)k_1}e^{\left(2\pi i\frac{|r_{-}|}{\ell_{\text{AdS}}}-\ell(p)\right)k_2}e^{-\Delta \ell(p)}\right)\;,
\end{equation}
with the holonomies introducing $\ell(p)$.

Indeed, the Wilson spool may be naturally understood as a sum over primitive geodesics, just like  $\log\zeta_{\Gamma}$. To see this, recall that the  Wilson spool (\ref{eq:spoolads}) is understood as an object that implements a sum over Wilson loops wrapping $n$ times around the holonomy cycle, here $\gamma_\phi$. For the BTZ black hole, this is particularly clear when we evaluate the contour integral, leaving:
\begin{align}
    \frac{1}{4}\mathbb{W}_j[a_L,a_R] 
    &= \sum_{n=1}^{\infty}\frac{1}{n}W_j[a_L,-a_R]^n~,
\end{align}
where $W_j$ is an ordinary Wilson loop associated to the representation. Implementing the path ordered exponentials (\ref{eq:holBTZ}) together with identification (\ref{hol}) realizes the Wilson spool as the sum 
\beq 
    \frac{1}{4}\mathbb{W}_j[a_L,a_R]= \frac{1}{2}\sum_{n=1}^{\infty}\frac{1}{n}\frac{e^{-n\ell(p)(\Delta-1)}}{\left(\cosh{(n\ell(p))}-\cos{(2\pi n|r_-|/\ell_{\text{AdS}})}\right)}~.
\label{eq:Wspoolbtzsimp2}\eeq
This geodesic formulation of the Wilson spool is consistent with the well-known saddle-point approximation for Wilson loops in higher-spin gravity, which isolates the dominant geodesic contribution: $W_{j}\sim e^{-\sqrt{c_j}\ell}$ at large $c_j$, where $c_j$ is the Casimir and $\ell$ is a geodesic length~\cite{Ammon:2013hba}.

\subsection{Spinning case}\label{2}

\noindent Let us now generalize the previous analysis to bosonic fields of arbitrary spin $s\in\mathbb{Z}$. 
In particular, we begin with the higher spin Wilson spool (\ref{eq:spooladsZ1}),
\begin{equation}\label{eq:s2}
    \log{Z_{\Delta,s}} = \frac{i}{4}\sum_{\mathcal{R}_{\Delta_s}^{\text{LW}}}\int_{\mathcal{C}}\frac{dz}{z}\frac{\cos{(z/2)}}{\sin{(z/2)}}\sum_{(\lambda_L,\lambda_R)}e^{iz(\lambda_L h_L-\lambda_R h_R)}~.
\end{equation}
The procedure for interchanging the infinite sums $(\lambda_L,\lambda_R)\in \mathcal{R}_{\Delta_s}^{\text{LW}}$ and contour integral works, \emph{mutatis mutandis}, as in the scalar field case.\footnote{The only difference is, in dilineating the infinite sums over $(\lambda_{L},\lambda_{R})$, we will have a remainder $\rho_{M}(z)$ term for both $(\lambda_L,\lambda_R)\in \mathsf{R}^{\text{LW}}_{j_L}\otimes\mathsf{R}^{\text{LW}}_{j_R}$ and $(\lambda_L,\lambda_R)\in \mathsf{R}^{\text{LW}}_{j_R}\otimes\mathsf{R}^{\text{LW}}_{j_L}$. For example, when $(\lambda_L,\lambda_R)\in \mathsf{R}^{\text{LW}}_{j_L}\otimes\mathsf{R}^{\text{LW}}_{j_R}$, the remainder $\rho_M(z)$ is (here we take $M=N$, without loss of generality)
$$\rho_M (z) = \frac{e^{-\frac{iz}{2}(h_L-h_R)}e^{iz(h_Lj_L-h_Rj_R)}}{4\sin{(\frac{z}{2}h_L)}\sin{(\frac{z}{2}h_R)}}\left(e^{-izMh_R}+e^{izMh_L}-e^{izM(h_L-h_R)}\right)\;.$$
As before in the scalar field case, $\rho_{M}(z)$ will not contribute any poles inside of the contour $\mathcal{C}$.
Further, with the holonomies (\ref{eq:btzhol}), and weights (\ref{eq:jljrAdS}), it is clear this term will vanish as $M\rightarrow\infty$. An analogous argument holds for $(\lambda_L,\lambda_R)\in \mathsf{R}^{\text{LW}}_{j_R}\otimes\mathsf{R}^{\text{LW}}_{j_L}$. }

Therefore, the 1-loop partition function becomes
\begin{align}\label{eq:hs1}
    \log{Z_{\Delta,s}} &= \frac{i}{4}\sum_{\mathcal{R}_{\Delta_s}^{\text{LW}}}\sum_{(\lambda_L,\lambda_R)}\int_{\mathcal{C}}\frac{dz}{z}\frac{\cos{(z/2)}}{\sin{(z/2)}}e^{iz(\lambda_L h_L-\lambda_R h_R)} \nonumber\\
    &= \log{\prod_{\mathcal{R}_{\Delta_s}^{\text{LW}}}\prod_{(\lambda_L,\lambda_R)}\left(1-e^{2\pi i((h_L \lambda_L -h_R \lambda_R))}\right)^{-1}}~.
\end{align}
Using the relation between the higher spin Selberg zeta function and $Z_{\Delta, s}$ from (\ref{eq:selr}) then allows us to identify the representation theory construction of the Selberg zeta function for arbitrary spin $s\in\mathbb{Z}$,
\begin{align}
    \zeta_{\Gamma}(\Delta_s) &= \prod_{\mathcal{R}_{\Delta_s}^{\text{LW}}}\prod_{(\lambda_L,\lambda_R)}\left(1-e^{2\pi i((h_L \lambda_L -h_R \lambda_R))}\right)\nonumber \\ 
    &= \prod_{\mathcal{R}_{\Delta_s}^{\text{LW}}}\prod_{(\lambda_L,\lambda_R)}\left<1-e^{2\pi  \partial_\phi} \right>~,
\end{align}
 generalizing \cite{Bagchi:2023ilg}. 

From this perspective, the factorization of the 1-loop determinant (\ref{eq:functionaldetAdS}) arises directly from the version of the spool \eqref{eq:spooladsZ1} that is expressed only using lowest weight representations. To see this, recall from \eqref{eq:reps2} that $\mathcal{R}_{\Delta_s}^{\text{LW}}$ includes representations with lowest weights given by either the pairs $(j_L,j_R)$ or $(j_R,j_L)$. Evaluating the product over these two possible representations gives
\begin{align}\label{eq:spinsel1}
    \zeta_{\Gamma}(\Delta_s) &= \prod_{\mathcal{R}_{\Delta_s}^{\text{LW}}}\prod_{(\lambda_L,\lambda_R)}\left<1-e^{2\pi  \partial_\phi} \right> \nonumber\\ 
    &=\prod_{k_1,k_2=0}^{\infty}\left(1-e^{2\pi i h_L k_1}e^{-2\pi ih_R k_2}e^{2(-a\Delta_{s}+ibs)}\right) \left(1-e^{2\pi i h_L k_1}e^{-2\pi ih_R k_2}e^{-2(a\Delta_{s}+ibs)}\right)\nonumber \\ 
    &= \zeta_{\mathbb{H}^{3}/\Gamma}\left(\Delta_{s}-\frac{isb}{a}\right)\cdot\zeta_{\mathbb{H}^{3}/\Gamma}\left(\Delta_{s}+\frac{isb}{a}\right).
\end{align}
This provides the lowest weight representation theory construction of the Selberg zeta function. However, because the Wilson spool can also be expressed using both lowest and highest weight representations in (\ref{eq:1loopWSde}), we are able to apply the formalism we have developed to reveal a further factorization of the Selberg zeta function,
\begin{equation}
    \zeta^2_{\Gamma} =\zeta_{\mathbb{H}^{3}/\Gamma}\left(\Delta_{s}^{\text{LW}}-\frac{isb}{a}\right)\zeta_{\mathbb{H}^{3}/\Gamma}\left(\Delta_{s}^{\text{LW}}+\frac{isb}{a}\right)\zeta_{\mathbb{H}^{3}/\Gamma}\left(-\Delta_{s}^{\text{HW}}+\frac{isb}{a}\right)\zeta_{\mathbb{H}^{3}/\Gamma}\left(-\Delta^{\text{HW}}_{s}-\frac{isb}{a}\right).
\end{equation}
Equating this to the square of (\ref{eq:spinsel1}) evaluated at $\Delta_s=\Delta_s^{\text{LW}}$ gives
\begin{equation}
    \zeta_{\mathbb{H}^{3}/\Gamma}\left(\Delta_{s}^{\text{HW}}-\frac{isb}{a}\right)\zeta_{\mathbb{H}^{3}/\Gamma}\left(\Delta_{s}^{\text{HW}}+\frac{isb}{a}\right) = \zeta_{\mathbb{H}^{3}/\Gamma}\left(\Delta_{s}^{\text{LW}}-\frac{isb}{a}\right)\zeta_{\mathbb{H}^{3}/\Gamma}\left(\Delta_{s}^{\text{LW}}+\frac{isb}{a}\right)\;.
\end{equation}
Thus, the Selberg zeta function can be expressed through either lowest or highest weight representations. This underscores what was found previously for the higher spin Wilson spool, this time throught the lens of the Selberg zeta function.

\section{Wilson spool from the $S^{3}$ trace formula}\label{SEC:DS}

\noindent Here we explicitly show how to recover the Wilson spool for three-dimensional Euclidean de Sitter space (EdS$_{3}$) via a trace formula for the sphere $S^{3}$. To this end, we begin by briefly reviewing the construction of the Wilson spool for $S^{3}$, focusing on the case of a massive scalar field (for higher-spin spools on EdS$_3$, see~\cite{Bourne:2024ded}). We will then show how analgous steps performed in Section~\ref{sec:BTZselbviaWS} to construct a Selberg-like zeta function from the spool 
fail in this case. Instead, we will construct a trace formula for $S^3$ and provide a new derivation of the $S^{3}$ Wilson spool from this trace formula. We also comment on how the Wilson spool itself can be considered a function that generalizes certain properties of the Selberg zeta function.

\subsection{Wilson spool for EdS$_3$: review}

\noindent Euclideanized de Sitter space has the geometry of a sphere. In three dimensions, the metric (in Hopf or torus coordinates) is
\begin{equation}\label{Eq:desittermeteric}    
\frac{ds^2}{\ell^{2}_{\text{dS}}}=\cos^2{(\rho)}\,d t_{E}^2+d\rho^2+\sin^2{(\rho)}\,d\phi^2~,
 \end{equation}
 where $\ell_{\text{dS}}$ is the curvature scale of dS$_{3}$, $\rho\in[0,\pi/2)$, and $\phi\in[0,2\pi)$.  The horizon of dS$_{3}$ in static patch coordinates (whose Lorentzian metric follows via the Wick rotation $t=-it_{E}$) is located at $\rho=\pi/2$, resulting in a conical singularity in the Euclidean geometry (\ref{Eq:desittermeteric}) that is removed by identifying $ t_{E}\sim t_{E}+2\pi n$, for integer $n$. The isometry group for $S^{3}$ is $SO(4)\simeq SU(2)_{L}\times SU(2)_{R}/\mathbb{Z}_{2}$, for left/right group actions $L/R$.

Accordingly, Euclidean dS$_{3}$ general relativity has a description in terms of a pair of $SU(2)$ Chern-Simons theories (see, e.g., Appendix A of \cite{Bourne:2024ded}). The background connections for the sphere  (\ref{Eq:desittermeteric}) are 
\begin{align}
    a_L &= i L_1 d\rho + i (\sin{\rho} L_2 - \cos{\rho} L_3)(d\phi - d t_{E})~,\nonumber \\
    a_R &= -i \bar{L}_1 d\rho - i(\sin{\rho} \bar{L}_2 + \cos{\rho} \bar{L}_3)(d\phi + d t_{E})~, 
\end{align}
where $\{L_{a}\}$ and $\{\bar{L}_{i}\}$ for $a=1,2,3$, denote the 
generators of the Lie algebras $\mathfrak{su}(2)_{L}$ and $\mathfrak{su}(2)_{R}$, respectively, obeying commutation relations $[L_{a},L_{b}]=i\epsilon_{ab}^{\quad c}L_{c}$, $[\bar{L}_{a},\bar{L}_{b}]=i\epsilon_{ab}^{\quad c}\bar{L}_{c}$, and $[L_{a},\bar{L}_{b}]=0$. These locally flat connections exhibit singularities when $\rho=0$ or $\rho=\pi/2$. Around these points, the connections have holonomies
\begin{equation}\label{eq:holdS}
\mathcal{P}e^{\oint_{\gamma}a_L} = g_L^{-1}e^{2\pi i h_L L_3}g_{L}~, \indent  \mathcal{P}e^{\oint_{\gamma}a_R} = g_R^{-1}e^{2\pi i h_R \bar{L}_3}g_{R}~,
\end{equation}
for periodic group elements $g_{L}=e^{iL_{1}\rho}$ and $g_{R}=e^{i\bar{L}_{1}\rho}$.  Specifically, for cycles $\gamma$ wrapping the horizon $\rho=\pi/2$, one has the holonomies
\begin{equation}
    h_L=1 \quad \text{and} \quad h_R =-1~.
\end{equation}
Meanwhile, cycles that wrap the singularity at the origin $\rho=0$ have $h_{L}=h_{R}=1$. 

In constructing the Wilson spool, it is necessary to specify representations that correspond to matter propagating on Wilson loops. Unlike for AdS$_3$, however, the EdS$_3$ spool for a massive scalar field is constructed from \emph{non-standard} representations that are distinct from the standard representations describing fields propagating on de Sitter (these were constructed specifically for the purposes of Chern-Simons theory in~\cite{Castro:2020smu,Castro:2023dxp}). Non-standard representations are highest-weight representations with highest weight $j$, but, unlike for the usual unitary $\mathfrak{su}(2)$ representations, $j$ can vary continuously. The Casimir is negative, indeed in terms of the mass $m$ of the scalar field it is
\begin{equation}
    j(j+1) = -\frac{1}{4}m^2 \ell_{\text{dS}}~,
\end{equation}
where $\ell_{\text{dS}}$ is the radius of curvature for the three-sphere. There are two relevant types of non-standard representations. First, there are \emph{complementary type} representations, which have
\be j = -\frac{1}{2}(1+\nu)~,\indent \bar{j}=j~,\indent\nu\in (-1,1)\;.\ee
These correspond to a light massive scalar with $m^2\ell_{\text{dS}} =1-\nu^2 <1$. Meanwhile, the \emph{principal type} representations are those where the highest weights take values
\be j = -\frac{1}{2}(1-i\mu)~,\indent \bar{j} = -1-j~, \indent \mu\in \mathbb{R}~,\ee
corresponding to a heavy massive scalar with $m^2 \ell_{\text{dS}}^2 = \mu^2+1 > 1$.
These non-standard representations admit characters that play an important role in the Wilson spool. For definiteness, we will focus on the principal series character, $\chi_j$, given by
\begin{equation} \label{eq:chars}
    \chi_{j}(x) = \sum_{k=0}^\infty e^{2\pi i  (j-k)x} = \frac{e^{2\pi i (j+1)x}}{e^{2\pi ix}-1}\;, \quad \chi_{j}(-x)=-\chi_{\bar{j}}(x)\;.
\end{equation}

Given two highest weight $\mathfrak{su}(2)$ representations, i.e., $\mathfrak{su}(2)_{L}$ and $\mathfrak{su}(2)_{R}$, their associated highest weights $j_L$ and $j_R$ are related to the conformal dimension and spin via\footnote{The conformal dimension and spin are eigenvalues of particular generators of the conformal algebra $\mathfrak{so}(1,3)\simeq \mathfrak{su}(2)_{L}\oplus\mathfrak{su}(2)_{R}$, the Lie algebra associated with the isometry group $SO(1,3)$ of Lorentzian dS$_{3}$. Consequently, unitary representations of $\mathfrak{so}(1,3)$ are labeled by $\Delta$ and $s$.}
\be \Delta = -j_L-j_R~, \indent s = j_L-j_R~.\ee
For the spinless case, which will be our focus, $j_L=j_R = j$ and $\Delta = -2j$. We will henceforth label these representations $\mathsf{R}_\Delta$, for either the principal or complementary type.
As in AdS$_{3}$, the shadow map $\Delta \rightarrow \bar{\Delta} = 2-\Delta$ leaves the Casimir invariant. Unlike for the BTZ black hole, however, where the shadow representations did not need to be considered due to boundary conditions, they must be included for Euclidean de Sitter \cite{Castro:2023dxp,Bourne:2024ded}.  The set of representations $\mathcal{R}_\Delta$ we include are therefore $ \mathcal{R}_{\triangle} = \{\mathsf{R}_{j}\otimes \mathsf{R}_{j},\mathsf{R}_{\bar{j}}\otimes \mathsf{R}_{\bar{j}}\}$.

We will not repeat the steps needed to derive the Wilson spool, and instead highlight elements relevant for the trace formula perspective (for a complete derivation, see~\cite{Castro:2023bvo,Bourne:2024ded}). As for the BTZ black hole, one follows a representation-theoretic DHS construction of the 1-loop determinant, e.g., for a real scalar field of mass $m$ (set by $\Delta$), 
\begin{equation}\label{eq:dsdet}
    Z_\Delta = \text{det}\left(-\nabla^2 + m^2\ell_{\text{dS}}^2\right)^{-1/2}~.
\end{equation}
Treat $Z_{\Delta}$ as a meromorphic function in $\Delta$. A necessary condition for a state $\ket{\psi}$ in representation $\mathcal{R}_{\Delta}$ to contribute a pole to $Z_{\Delta}$ is that $|\psi\rangle$ satisfy the mass shell condition
\begin{equation}\label{eq:massshellds}
    (-\nabla^2+m^2\ell_{\text{dS}}^2)\ket{\psi} = (2c_{2,L}+2c_{2,R}+m^2\ell_{\text{dS}}^2)\ket{\psi}=0~.
\end{equation}
Without analytically continuing the mass, this corresponds precisely to highest weight, non-standard representations, with the highest weight state labeled by $j<0$ satisfying
\begin{equation}\label{eq:msquared}
    j(j+1) = -\frac{1}{4}m^2 \ell_{\text{dS}}^2~.
\end{equation}
Once analytically continued, this also captures the standard $\mathfrak{su}(2)$ representations with positive Casimir. As emphasized in~\cite{Bourne:2024ded}, an additional condition of global hyperbolicity in fact excludes the non-standard representations. As such, the pole structure of $Z_{\Delta}$ is set entirely by the standard representations.

Following a similar set of steps laid out in Section \ref{sec:wsads}, the 1-loop partition function for massive scalar field on $S^{3}$ is 
\be \log{Z_\Delta} = \frac{1}{4}\mathbb{W}_j~,\label{eq:WilspoolS3Z1}\ee
for Wilson spool
\begin{align}\label{eq:spoolds}
    \mathbb{W}_j [a_{L},a_{R}] &= i\int_{\mathcal{C}}\frac{dz}{z}\frac{\cos{(z/2)}}{\sin{(z/2)}}\text{Tr}_{R_j}\left(\mathcal{P}e^{\frac{z}{2\pi}\oint a_L}\right)\text{Tr}_{R_j}\left(\mathcal{P}e^{-\frac{z}{2\pi}\oint a_R}\right) \nonumber\\ 
    &= i\int_{\mathcal{C}}\frac{dz}{z}\frac{\cos{(z/2)}}{\sin{(z/2)}}\chi_j\left(\frac{z}{2\pi}h_{L}\right)\chi_{j}\left(-\frac{z}{2\pi}h_{R}\right)\nonumber \\ 
    &= -\frac{i}{4}\int_{\mathcal{C}}\frac{dz}{z}\frac{\cos{(z/2)}}{\sin^3{(z/2)}}e^{iz(2j+1)}~.
\end{align}
Here the contour $\mathcal{C}$ is negatively oriented, wrapping once around $z=0$ and twice around the positive real axis, as depicted in Figure~\ref{fig:Contour2}. 
This will figure later in our discussion about the trace formula.

\subsection{Selberg zeta function for $S^3$?}

\noindent The construction of the Selberg zeta function relies on the existence of nontrivial cycles with associated primitive geodesics, which exist for discrete quotient manifolds, such as $\mathbb{H}^n/\Gamma$. Since $S^3$ is not a discrete quotient of another manifold, one should not expect to successfully define a meaningful Selberg-like zeta function for the sphere. However, since $S^3$ is a special case of lens spaces $S^3/\Gamma$ (with $\Gamma$ a discrete subgroup of $SO(4)$), one can naively seek a Selberg-like zeta function for $S^3$, in analogy to the Wilson spool analysis for Euclidean BTZ in Section \ref{sec:BTZselbviaWS}. Indeed, the Wilson spool was explicitly constructed for $S^3$ using the fact that, while geodesics on the sphere are contractible to a point, $S^3$ admits non-trivial holonomies~\cite{Castro:2023dxp,Castro:2023bvo,Bourne:2024ded}. Thus, despite the lack of nontrivial primitive geodesics, it is natural to wonder if a representation-theoretic Selberg zeta function could be constructed from the Wilson spool for the sphere, as outlined in Section~\ref{sec:BTZselbviaWS} for the BTZ black hole.

It turns out this method for deriving a Selberg zeta function for the sphere is not possible. Recall that the derivation presented in Section \ref{sec:BTZselbviaWS}  involved a simple interchange of an infinite sum (corresponding to the trace over states in the representation) with the contour integral defining the spool. We carefully demonstrated this operation was allowed by splitting the sum into a finite piece and its remainder, $\rho_{M,N}(z)$. Attempting this for the EdS$_3$ spool, however, we find\footnote{Without loss of generality, we can let $M=N$ to write $\rho_M(z)=-\frac{e^{iz(2j-2M+1)}}{4\sin^2{(z/2)}}(1+4i\sin{(\frac{Mz}{2})}e^{\frac{iz}{2}M})$.}
\begin{equation}
    \lim_{M,N\rightarrow\infty}\int_{\mathcal{C}}\frac{dz}{z}\frac{\cos{(z/2)}}{\sin{(z/2)}}\rho_{M,N}(z) =\infty~.
\end{equation}
Thus, the remainder term diverges and interchanging the sum and integral is not allowed. 

Amusingly, had we naively interchanged the sum and the integral in (\ref{eq:spoolds}) while dropping the remainder $\rho_{M,N}(z)$ as well as the zero mode contribution, then 
\begin{align}
    \frac{i}{4}\sum_{k_1=0}^{\infty}\sum_{k_2=0}^{\infty}\int_{\mathcal{C}/\{0\}}\frac{dz}{z}\frac{\cos{(z/2)}}{\sin{(z/2)}}e^{iz(2j-k_1-k_2)} &= -\sum_{k_1=0}^{\infty}\sum_{k_2=0}^{\infty}\log{(1-e^{2\pi i(2j-k_1-k_2)})}~.
\end{align}
Introducing $\zeta_{S^{3}}$ through $\mathbb{W}_j/4 =-\log{\zeta_{S^{3}}}$, and expressing the product in terms of the generator of rotations along the angle corresponding to Euclidean time translations,
\be i \partial_{\tau} = L_0 + \bar{L}_0~, \ee
we obtain 
\begin{align}\label{eq:repsel}
    \zeta_{S^{3}}(j) &= \prod_{k_1=0}^{\infty}\prod_{k_2=0}^{\infty}\left(1-e^{2\pi i(2j-k_1-k_2)}\right)\nonumber \\  &= \prod_{\text{descendants}}\left<1-e^{-2\pi \partial_{\tau}}\right>~,
\end{align}
where the Euler-like product is taken over descendants within the representation $\mathsf{R}_{\Delta}$ associated to the Wilson spool. While $\zeta_{S^{3}}$ has a similar form to what would be obtained using the representation theory construction, it is deceptive since the overall $k_1,k_2$ dependence drops out, leading to a divergent infinite product.

While the Wilson spool no longer matches the (log of) the Selberg zeta function in Euclidean de Sitter as it does for BTZ, it retains some features reminiscent of the Selberg zeta function.
Indeed, the non-standard representations for the EdS$_3$ spool figure in a representation-theoretic expression for a sum over holonomies (the generalization of geodesics), just with the integral and sum left uninterchanged---an ``Euler-like'' product. Further, 
the poles of the spool lie precisely at the \emph{standard} $\mathfrak{su}(2)$ representations, in analogy with the zeros of the Selberg zeta function.  
Additionally, just as the Selberg zeta function arose from a trace formula \cite{selberg1956harmonic,Voros:1986vw}, 
we will now demonstrate a trace formula for $S^{3}$ gives rise to the Wilson spool.

\subsection{Trace formula for $S^{3}$}\label{sec:traceformS3}

\noindent First we will construct a trace formula for $S^{3}$. Broadly speaking, a trace formula is a  between spectral data of particular operators and geometric data, i.e., integrals along orbits on a particular space. A simple example of a trace formula is the Poisson resummation formula
\begin{equation}\label{eq:pois}
    \sum_{m\in\mathbb{Z}}h(m)=\sum_{n\in\mathbb{Z}}\int_{\mathbb{R}}h(z)e^{2\pi inz}dz~,
\end{equation}
for a sufficiently smooth test function $h\in L^{1}(\mathbb{R})$. A relevant application of the Poisson resummation formula (\ref{eq:pois}) is that it can be used to re-express the spectral data of a Laplacian of a massive particle confined to a circle $S^{1}\simeq\mathbb{R}/\mathbb{Z}$. 

To see this, note the energy eigenvalue problem $\nabla^{2}\psi_{n}=E_{n}\psi_{n}$ for a particle of unit mass on $S^{1}$ of length $2\pi$ has the kinetic operator $(-\nabla^{2})=\partial_{x}^{2}/2$ with eigenfunctions $\psi_{n}=\frac{1}{\sqrt{2\pi}}e^{-inx}$ and energies $E_{n}=n^{2}/2$. 
The associated heat kernel has the spectral decomposition, $K^{S^{1}}(x,y;2\pi)=\sum_{n}\psi_{n}(x)\psi_{n}^{\ast}(y)e^{-2\pi E_{n}}$, such that 
\beq \text{tr}(e^{-2\pi\nabla^{2}})=\int_{S^{1}}K^{S^{1}}(x,x;2\pi)dx=\sum_{n\in\mathbb{Z}}e^{-2\pi E_{n}}\;,\label{eq:spectracS1}\eeq
which we can interpret as the canonical partition function for the system with (inverse) temperature $2\pi$. Applying the Poisson resummation formula (\ref{eq:pois}) with test function $h(z) = e^{-2\pi E_z}$, one may write 
\beq \text{tr}(e^{-2\pi \nabla^{2}})=\sum_{n\in\mathbb{Z}}\int_{\mathbb{R}}e^{-2\pi E_{z}}e^{2\pi inz}dz\;.\eeq
Thus, the Poisson resummation formula relates the trace of a kinetic operator to the sum over the periodic orbits of length $2\pi|n|$ on $S^{1}$. The above is easily generalizable for a circle of length $\beta$ (merely replace all factors of $2\pi$ for $\beta$), and the functional determinant for the kinetic operator is cast in terms of the trace formula,
\beq \log \text{det}(-\nabla^{2})=-\int_{0}^{\infty}\frac{d\beta}{\beta}\text{tr}(e^{-\beta \nabla^{2}})\;,\label{eq:heatkernelexample}\eeq
where to obtain this expression we have evaluated the one-loop determinant using the heat kernel trace and again applied the Poisson resummation formula.\footnote{As an alternate derivation that sheds light on the connection to the Wilson spool, make use of the identity $\log{\det} = \tr \log $ combined with the Schwinger trick, \eqref{eq:Schwinger}, which was used to construct the Wilson spool. Deriving \eqref{eq:heatkernelexample} this way requires interchanging a trace with the integral over the Schwinger parameter, which we may be invalid for certain cases of interest. One can view our technical formalism using Fredholm determinants as a precise way of dealing with such subtleties. We thank the anonymous referee for this observation.} In general, for a free particle on a symmetric space, Poisson resummation relates the spectral sum of Laplacians to a sum over geodesics on a maximal torus, which has proven useful to show equivalence for various representations of the heat kernel (cf. Chapter 6 of \cite{Camporesi:1990wm}). Analogously,  the Selberg trace formula is viewed as a generalization of the Poisson resummation formula for discrete quotients of hyperbolic space (see, e.g.,  \cite{Balazs:1986uj,Voros:1986vw,Bytsenko:1994bc,marklof2004selberg} for pedagogical treatments).

Spectral trace formulae of the type (\ref{eq:spectracS1}) are well known for higher-dimensional spheres and their lens space quotients; indeed they have been used to evaluate 1-loop partition functions (see, e.g., \cite{David:2009xg}). Here, following the spirit of Voros' work on zeta function factorization \cite{Voros:1986vw}, we construct a different trace formula for a scalar field on the 3-sphere based on the Fredholm determinant.\footnote{Voros' work on zeta function factorizations was extended to $n$-dimensional spheres by Choi and Quine \cite{10.1216/rmjm/1181072081}, and used to compute spherical functional determinants. As far as we can tell, the trace formula we construct is new to the literature.} More specifically, our focus will be to find an appropriate test function, different than that used for generalizations of (\ref{eq:spectracS1}), which gives rise to a Hadamard factorization of the Fredholm determinant for a massive scalar on the 3-sphere. From this test function we will be able to directly construct the Wilson spool from a trace formula.

\subsection*{The Fredholm determinant and its trace formula}

\noindent Consider a sequence $\{\lambda_{\ell}\}$ of non-zero complex numbers (note that zero modes have to be dealt with separately). For our purposes, we consider the eigen-spectrum of a kinetic differential operator $\hat{\mathcal{O}}$ defined on a compact manifold by
\begin{equation}\label{eq:Oform}
    \hat{\mathcal{O}} =-\nabla^2 -\lambda~,
\end{equation}
where $\{\lambda_\ell\}$ denotes the eigenvalues of the Laplacian, and $\lambda$ is a real spectral parameter.
A particular spectral function of interest is the Fredholm determinant associated to the Laplacian on the $3$-sphere or lens space, cast here as a canonically regularized infinite product\footnote{For $\lambda_{\ell}$ being eigenvalues of an elliptic (pseudo-) differential operator of order $m$ on a $d$-dimensional (compact) manifold, the sequence $\{\lambda_{\ell}\}$ has order $\mu\equiv d/m$, and the infinite product without the exponential factor only converges for sequences with $\mu<1$. Presently, $\mu>1$ and thus we use the Weierstrass canonical product, which introduces the exponential factor in (\ref{eq:fred}) \cite{Voros:1986vw}.}
\begin{equation}\label{eq:fred}
    \mathbf{\Delta} (\lambda)\equiv \text{det}(\mathbbm{1}+\lambda\left(\nabla^2\right)^{-1}) =\prod_{\ell=0}^{\infty}\left(1-\frac{\lambda}{\lambda_\ell}\right)^{d_\ell}e^{\frac{\lambda}{\lambda_\ell}d_{\ell}}\;,
\end{equation}
where $d_{\ell}$ are degeneracies of eigenvalues $\lambda_{\ell}$. The Fredholm determinant thus generalizes the determinant of a finite linear operator, and has featured in various contexts ranging from nuclear physics, random matrices, to 2D quantum gravity, cf. \cite{Wheeler:1937zz,Forrester+2010,Johnson:2021zuo}. Pertinently, $\mathbf{\Delta}(\lambda)$ is related 
to 
the functional determinant  $D(\lambda)\equiv \text{det}(\hat{\mathcal{O}})=\prod_{\ell}(\lambda_{\ell}-\lambda)^{d_{\ell}}$ via\footnote{Indeed, $\mathbf{\Delta}(\lambda)=\left(\prod_{\ell}(\lambda_{\ell}-\lambda)^{d_{\ell}}\right)\left(\prod_{\ell}\lambda_{\ell}^{d_{\ell}}\right)\left(\prod_{\ell}e^{\frac{\lambda}{\lambda_{\ell}}\lambda}\right)=D(\lambda)e^{\zeta'_{\hat{\mathcal{O}}}(0)}e^{\lambda \zeta_{\hat{\mathcal{O}}}(1)}$. For a more complete derivation taking into account an appropriate regularization, see Appendix~\ref{app:genzetafuncs}.}
\beq D(\lambda)=\mathbf{\Delta}(\lambda)e^{-\zeta'_{\hat{\mathcal{O}}}(0)}e^{-\lambda\zeta_{\hat{\mathcal{O}}}(1)}\;,\label{eq:22}\eeq
where $\zeta_{\hat{\mathcal{O}}}(s)=\sum_{\ell=0}^{\infty}d_{\ell}\lambda_{\ell}^{-s}$ denotes the spectral zeta function, with $\zeta'_{\hat{\mathcal{O}}}(s)\equiv\partial_{s}\zeta_{\hat{\mathcal{O}}}(s)$. Since we are focusing on scalar fields, the functional determinant is related to the 1-loop partition function via $D(\lambda)^{-1/2}=Z_{\Delta}$. Further, taking the logarithm and differentiating twice with respect to $\lambda$ gives $\partial^{2}_{\lambda}\log(D(\lambda))=\partial_{\lambda}^{2}\log(\mathbf{\Delta}(\lambda))$. Since the Fredholm determinant is canonically regularized, we are guaranteed to be able to pass derivatives through the infinite sums appearing in $\log(D(\lambda))$. This fact is implicitly used when applying the Poisson resummation formula below. 
\par
From the Fredholm determinant~\eqref{eq:fred}, one can construct the trace of the resolvent, $R(\lambda)$, equal to the logarithmic derivative of $\mathbf{\Delta}(\lambda)$,
\begin{equation}
    R(\lambda) \equiv -\frac{d}{d\lambda}\log{\mathbf{\Delta}(\lambda)}=\sum_{\ell=0}^{\infty}\frac{d_{\ell}\lambda}{\lambda_{\ell}(\lambda_{\ell}-\lambda)}~. \label{eq:resolvent}
\end{equation}
In particular, we are interested in the derivative of the resolvent, 
\begin{align}\label{eq:Rpgen}
    R'(\lambda) 
    &= \sum_{\ell=0}^{\infty}\frac{d_\ell}{(\lambda-\lambda_\ell)^2} ~.
\end{align}
Let us now apply the Poisson summation formula for the resolvent of the Fredholm determinant. To this end, consider a test function $h(z)$ with an associated degeneracy function $d(z)$, such that the combination $d(z)h(z)\in L^1(\mathbb{R})$, and is even in $z$. We pick the test function and degeneracy function to be one appropriate for expressing the resolvent as a sum over $\mathbb{Z}$, resulting in:
\beq R'(\lambda)=\frac{1}{2}\sum_{n\in\mathbb{Z}}d(n)h(n)=\frac{1}{2}\sum_{n\in\mathbb{Z}}\int^{\infty}_{-\infty}d(z)h(z)e^{2\pi i nz}dz\;,\label{eq:traceform}\eeq
where we used the Poisson resummation formula (\ref{eq:pois}) to arrive at the second equality. We will leave this test function unspecified for now, to account for any Fredholm determinant of the form \eqref{eq:fred}; in the next section, we will give an explicit form of this test function for the specific case of the three-sphere (see Appendix~\ref{app:traceformbeyondS3} for the lens space quotient of the 3-sphere). 

\begin{figure}[t]
    \centering
    \resizebox{0.5\linewidth}{!}{
    \begin{tikzpicture}[scale=1.1, decoration={
        markings,
        mark=between positions 0.05 and 0.95 step 2cm with {\arrow[line width=2pt]{>}}
    }]
        \draw[help lines,<->,line width=2pt] (-6.5,0) -- (6.5,0);
        \draw[help lines,<->,line width=2pt] (0,-4.5) -- (0,4.5);

        \draw[black, line width=1.5pt, postaction=decorate] 
            (6.5,-1.0) -- (2,-1.0)
            arc (270:90:1.0)
            -- (6.5,1.0);
        \node[below] at (7.5,.43) {\LARGE{$\text{Re}(z)$}};
        \node[above] at (0,4.5) {\LARGE{$\text{Im}(z)$}};
        \node at (1.6,1.5) {\LARGE{$\Gamma$}};

        \foreach \x in {-6,-4.5,-3,-1.5,0,1.85,3,4.5,6} {
            \filldraw[violet] (\x,0) circle (2.5pt);
        }
    \end{tikzpicture}
    }
    \caption{The contour $\Gamma$ wrapping around the real axis enclosing poles  at $z=n$ for $n\in\mathbb{N}$.}
    \label{fig:Contour2}
\end{figure}
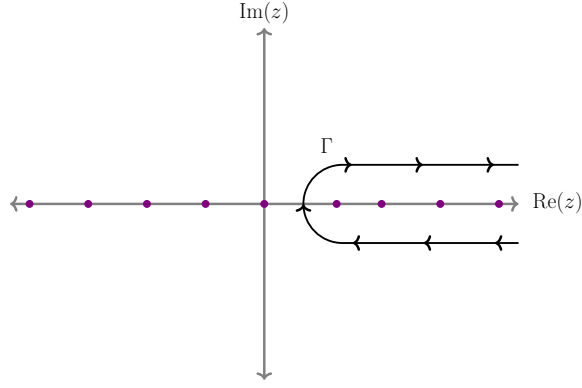

We can interchange the infinite sum and integral with an appropriate $i\epsilon$ prescription to ensure convergence. More carefully, using the even parity of $d(z)h(z)$ to split the integral over $\mathbb{R}$ into two integrals over $\mathbb{R}^{+}$, and rewriting the sum over $n$ into a sum over $|n|$ gives
\beq 
\begin{split}
 R'(\lambda)&=\frac{1}{2}\sum_{n\in\mathbb{Z}}\int_{0}^{\infty}d(z)h(z)e^{2\pi i |n|z}dz+\frac{1}{2}\sum_{n\in\mathbb{Z}}\int_{0}^{\infty}d(z)h(z)e^{-2\pi i |n|z}dz\;.
\end{split}
\eeq
We can then interchange the sum and integral in the first (second) term by deforming $z\to z+i\epsilon$ ($z\to z-i\epsilon$). In so doing, 
\begin{align}
    R'(\lambda)
    &= \frac{i}{2}\lim_{\epsilon\rightarrow0}\left(\int_{0 +i\epsilon}^{\infty+i\epsilon}+\int_{\infty-i\epsilon}^{0-i\epsilon}\right)d(z)h(z)\cot{(\pi z)}dz\;.
\end{align}
 Assuming the integrand does not have any poles at $z=0$, we may express the two integrals as a single integral over a contour $\Gamma$, such that 
\begin{equation}\label{eq:trS3}
  R'(\lambda)= \frac{1}{2}\sum_{n\in\mathbb{Z}}d(n)h(n)
    = \frac{i}{2}\int_{\Gamma}d(z)h(z)\cot{(\pi z)}dz\;,
\end{equation}
where $\Gamma$ is negatively oriented and wraps around the real axis, enclosing poles at $z=n$ for $n\in\mathbb{N}$ (see Figure \ref{fig:Contour2}). Note that while we uncovered this trace formula with the resolvent $R'(\lambda)$ in mind, the preceding derivation in fact holds for more general smooth test functions.\footnote{If the integrand has removable singularities, the Poisson summation formula may be used by introducing a shift $i\delta$ in a way that removes the singularities. The trace formula (\ref{eq:trS3}) can then be evaluated by taking $\delta\rightarrow 0$, omitting the removable singularity when summing the residues.}

From the relation (\ref{eq:22}), the functional determinant $D(\lambda)$ can be directly obtained from the resolvent operator $R'(\lambda)$ by an appropriate double integration
\begin{equation}\label{eq:intd}
    \log{D(\lambda)} = -\int_{-\infty}^{\lambda}\int_{-\infty}^{\lambda'}d\lambda'd\lambda'' R'(\lambda'')~.
\end{equation}
We will use this relation below when we construct the spool.

\subsection{New derivation of the Wilson spool}\label{sec:4.2}
\noindent The Wilson spool for Euclidean de Sitter can be derived from the resolvent $R'(\lambda)$. 
Let $\hat{\mathcal{O}}$ be the massive scalar Laplacian on $S^{3}$, 
\begin{equation}
    \hat{\mathcal{O}} =-\nabla^2 +m^2~.
\end{equation}
Comparing to \eqref{eq:Oform} and \eqref{eq:msquared}, we make the identification
\be \label{eq:lambdatoj}
\lambda=-m^2~, \indent j=-\frac{1}{2}\left(1\pm \sqrt{1+\lambda}\right)\equiv-\frac{1}{2}\Delta_{\pm}~.
\ee
The massless scalar Laplacian $-\nabla^2$ on the three-sphere has eigenvalues and degeneracies
\begin{equation}
\lambda_{\ell}=\ell(\ell+2)~, \indent d_{\ell}=(\ell+1)^{2}~,
\end{equation}
for positive integer $\ell$. \par
For the purposes of deriving the Wilson spool, it will be useful to also define a Hadamard factorization of the Fredholm determinant. Specifically, we define
\begin{equation}
    \mathbf{\Delta}^{+} (z)=\prod_{\ell=0}^{\infty}\left(1-\frac{z}{\ell+1}\right)^{d_{\ell}}e^{d_{\ell}\left(\frac{z}{\ell+1}+\frac{z^2}{2(\ell+1)^2}\right)},
\end{equation}
which factorizes $\mathbf{\Delta} (\lambda)$ up to an entire function $Q(\ell)$ that vanishes when taking the derivatives with respect to $\lambda$:
\begin{equation}
\mathbf{\Delta}(\lambda)=Q(\ell)\prod_{\ell=0}^{\infty}\left(1-\frac{1+\lambda}{(\ell+1)^2}\right)^{d_\ell}e^{\frac{d_\ell}{(\ell+1)^2}(\lambda+1)}=Q(\ell)\mathbf{\Delta}^{+}(\sqrt{1+\lambda})\mathbf{\Delta}^{+}(-\sqrt{1+\lambda})~.\label{eq:factorizedDelta}
\end{equation}
Importantly, this is expressed in terms of the squares $(\ell+1)^2$, which can be shifted to create an even test function, rather than the eigenvalues $\{\lambda_\ell\}$.
\par
From \eqref{eq:Rpgen}, the derivative of the resolvent for $S^3$ is given by
\begin{align}\label{eq:Rp}
    R'(\lambda) 
   = \frac{1}{2}\sum_{n\in\mathbb{Z}}\frac{n^2}{(1+\lambda-n^2)^2}\;.
\end{align}
Thus, comparing to~\eqref{eq:traceform} the appropriate test function is
\beq h(n)=\frac{1}{(1+\lambda-n^{2})^{2}}~,\label{eq:testfuncRp}\eeq
with degeneracy function $d(n)=n^2$. For the derivation of the spool, it will be useful to decompose this test function into two terms that combine to give the full range of integration. We can find the canonical separation using the Hadamard factorization of the Fredholm determinant, \eqref{eq:factorizedDelta}. Taking the logarithmic derivative of $\mathbf{\Delta}(\lambda)$ expressed in this factorized form, we obtain four terms, two from each of $\mathbf{\Delta}^+(\pm \sqrt{1+\lambda})$. The contributions from the exponential parts cancel out, leaving 
\be 
 R'(\lambda)=\frac{1}{2}\sum_{n\in\mathbb{Z}}n^{2}h(n)~,
\ee
with the factorized test function given by
\begin{equation}
    h(n)
    = \frac{2\sqrt{1+\lambda}-n}{4\left(1+\lambda\right)^{3/2}(n-\sqrt{1+\lambda})^2}+\frac{2\sqrt{1+\lambda}+n}{4\left(1+\lambda\right)^{3/2}(n+\sqrt{1+\lambda})^2}~.
\end{equation}
This factorization of $\lambda$ corresponds to a factorization of both roots $\Delta_{\pm}$ of $j$ given in~\eqref{eq:lambdatoj}.\footnote{The factorization in terms of $\Delta_{\pm}$ is, $h(n)=\frac{-2(1-\Delta_+)-n}{4(1+\lambda)^{3/2}(n+1-\Delta_+)^2}+\frac{2(1-\Delta_-)+n}{4(1+\lambda)^{3/2}(n+1-\Delta_-)^2}$~.} For de Sitter, unlike for AdS, there are no boundary condition forbidding the $\Delta_-$ solution, and both representations can have important physical applications (see, e.g., \cite{Castro:2020smu, Xiao:2014uea}). 
\par
\begin{figure}[t]
    \centering
    \resizebox{0.9\linewidth}{!}{ 
    \begin{tikzpicture}[scale=1.1, decoration={markings,
        mark=at position 0.5cm with {\arrow[line width=2.5pt]{>}},
        mark=at position 2cm with {\arrow[line width=2.5pt]{>}},
        mark=at position 3.5cm with {\arrow[line width=2.5pt]{>}},
        mark=at position 5.5cm with {\arrow[line width=2.5pt]{>}},
        mark=at position 7.85cm with {\arrow[line width=2.5pt]{>}},
        mark=at position 9cm with {\arrow[line width=2.5pt]{>}},
        mark=at position 11cm with {\arrow[line width=2.5pt]{>}}
    }]
        \draw[help lines,<->,line width=2pt] (-6.5,0) -- (6.5,0) coordinate (xaxis);
        \draw[help lines,<->,line width=2pt] (0,-4.5) -- (0,4.5) coordinate (yaxis);
        \path[draw,line width=2pt,postaction=decorate, black] (-.75,-4.5)  -- (-.75,4.5) ;
        \path[draw,line width=2pt,postaction=decorate, black] (.75,-4.5)  -- (.75,4.5) ;
        \node[below] at (7.5,.43) {\huge{$\text{Re}(z)$}};
        \node[above] at (0,4.5) {\huge{$\text{Im}(z)$}};
        \node at (1.5,2) {\huge{$\mathcal{C}$}};
        \filldraw[violet] (-6,0) circle (2pt) ;
        \filldraw[violet] (-4.5,0) circle (2pt) ;
        \filldraw[violet] (-3,0) circle (2pt) ;
        \filldraw[violet] (-1.5,0) circle (2pt) ;
        \filldraw[violet] (0,0) circle (2pt) ;
        \filldraw[violet] (1.5,0) circle (2pt) ;
        \filldraw[violet] (3,0) circle (2pt) ;
        \filldraw[violet] (4.5,0) circle (2pt) ;
        \filldraw[violet] (6,0) circle (2pt) ;
        \node at (0,-5.2) {\huge{(a)}};
    \end{tikzpicture}
    \hspace{1cm}
    \begin{tikzpicture}[scale=1.1, decoration={markings,
        mark=at position 0.5cm with {\arrow[line width=2.5pt]{>}},
        mark=at position 2cm with {\arrow[line width=2.5pt]{>}},
        mark=at position 3.5cm with {\arrow[line width=2.5pt]{>}},
        mark=at position 5.5cm with {\arrow[line width=2.5pt]{>}},
        mark=at position 7.85cm with {\arrow[line width=2.5pt]{>}},
        mark=at position 9cm with {\arrow[line width=2.5pt]{>}},
        mark=at position 11cm with {\arrow[line width=2.5pt]{>}}
    }]
        \draw[help lines,<->, line width=2pt] (-6.5,0) -- (6.5,0) coordinate (xaxis);
        \draw[help lines,<->,line width=2pt] (0,-4.5) -- (0,4.5) coordinate (yaxis);
        \path[draw,line width=1.5pt,postaction=decorate, black] (6.5,-1) -- (2,-1)  arc (-100,1)(270:90:1) -- (6.5,1) ;
        \path[draw,line width=1.5pt,postaction=decorate, black] (6.5,-.5) -- (1.91,-.5)  arc (-100,1)(270:90:.5) -- (6.5,.5) ;
        \path[draw,line width=1.5pt,postaction=decorate, black]  (.75,0)  arc (-100,1)(0:-375:.75) ;
        \node[below] at (7.5,.43) {\huge{$\text{Re}(z)$}};
        \node[above] at (0,4.5) {\huge{$\text{Im}(z)$}};
        \node at (1.5,2) {\huge{$\mathcal{C}$}};
        \filldraw[violet] (-6,0) circle (2pt) ;
        \filldraw[violet] (-4.5,0) circle (2pt) ;
        \filldraw[violet] (-3,0) circle (2pt) ;
        \filldraw[violet] (-1.5,0) circle (2pt) ;
        \filldraw[violet] (0,0) circle (2pt) ;
        \filldraw[violet] (1.85,0) circle (2pt) ;
        \filldraw[violet] (3,0) circle (2pt) ;
        \filldraw[violet] (4.5,0) circle (2pt) ;
        \filldraw[violet] (6,0) circle (2pt) ;
        \node at (0,-5.2) {\huge{(b)}};
    \end{tikzpicture}
    } 
    
    \caption{(a) The contour $\mathcal{C}$ for the Euclidean de-Sitter Wilson spool. (b) The deformed contour $\mathcal{C}$.}
    \label{fig:Contour3}
\end{figure}
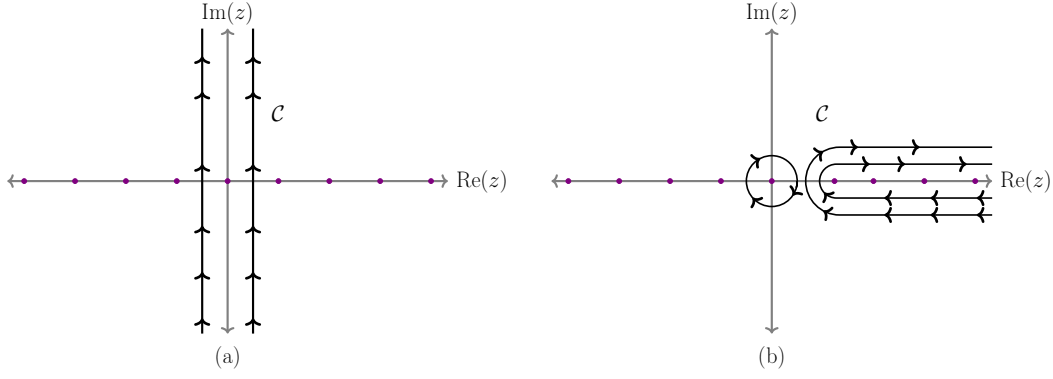

Inserting the factorized test function into the trace formula (\ref{eq:trS3}) and integrating twice using \eqref{eq:intd} gives
\begin{align}\label{eq:trw}
    \log{D(\lambda)} 
    = &-\frac{i}{8}\int_{-\infty}^{\lambda}\int_{-\infty}^{\lambda'}d\lambda'd\lambda'' \int_{\Gamma}\frac{z^2\left(2\sqrt{1+\lambda ''}-z\right)}{\left(1+\lambda ''\right)^{3/2}(z-\sqrt{1+\lambda ''})^2}\cot{(\pi z)}dz \nonumber  \\ 
    &-\frac{i}{8}\int_{-\infty}^{\lambda}\int_{-\infty}^{\lambda'}d\lambda'd\lambda'' \int_{\Gamma}\frac{z^2\left(2\sqrt{1+\lambda ''}+z\right)}{\left(1+\lambda ''\right)^{3/2}(z+\sqrt{1+\lambda ''})^2}\cot{(\pi z)}dz~.
\end{align}
We would like to interchange the $\lambda$ integrals with the contour integral over $\Gamma$. Because the pole structure of the integrand depends on $\lambda$, however, such an interchange is not allowed. To circumvent this, we re-express the function $h(z)$ using the following Laplace transform, 
\begin{equation}\label{eq:lapl}
    \mathcal{L}\Bigl\{\left(t\sqrt{1+\lambda}\pm 1\right)e^{\mp t\sqrt{1+\lambda}}\Bigr\}(z) = \int_{0}^{\infty}\left(t\sqrt{1+\lambda}\pm 1\right)e^{-t(z\pm \sqrt{1+\lambda})} dt = \frac{2\sqrt{1+\lambda}\pm z}{(z\pm\sqrt{1+\lambda})^2}~.
\end{equation} 
Since the bounds of integration for $\lambda$ extends from $-\infty$ to $\lambda$, the presence of $\sqrt{1 + \lambda}$ introduces a branch point at $\lambda = -1$. To handle this singularity, we employ a shift $t \rightarrow t \pm i\epsilon$ such that (\ref{eq:lapl}) converges for $t > 0$. We can insert the Laplace transformation (\ref{eq:lapl}) with the appropriate contour shift into (\ref{eq:trw}), then interchange and evaluate the $\lambda$ integrals to obtain:\footnote{In the limit $\lambda \rightarrow -\infty$,  convergence is determined by the exponential term. Taking the limit $\lambda \rightarrow -\infty$ is equivalent to taking $\sqrt{1 + \lambda} \rightarrow i\infty$. For the `$+$' term, we shift the contour to $t \rightarrow t - i\epsilon$ to ensure the integrand converges in the limit $\lambda \rightarrow -\infty$ for $t>0$. For the `$-$' term, we shift the contour by $t \rightarrow t + i\epsilon$. A hard cutoff at $t = 0$ regulates the $t = 0$ divergence that appears after integrating with respect to $\lambda$.}
\begin{align}\label{eq:lamshift}
    \log{D(\lambda)}=-\frac{i}{2}\sum_{\pm}\int_{\cross\pm i\epsilon}^{\infty\pm i\epsilon}\frac{dt}{t}\int_{\Gamma}z^2\cot{(\pi z)}e^{-t(z\mp\sqrt{1+\lambda})}dz.
\end{align}
After the $\lambda$ integrals are performed, we can take the limit $\epsilon\rightarrow 0$. Now we must regulate the $t=0$ divergence. This can be done by expanding (\ref{eq:lamshift}), flipping the bounds of integration and taking both $t\rightarrow -t$ and $z\rightarrow -z$ for one of the terms to obtain
\begin{align}\label{eq:W}
    \log{D(\lambda)}  
    &=- \frac{i}{2}\left[\int_{-\infty}^{-\cross}\frac{dt}{t}+\int_{\cross}^{\infty}\frac{dt}{t}\right]\int_{\Gamma} z^2\cot{(\pi z)}e^{-t(z-\sqrt{1+\lambda})}dz~.
\end{align}
The $i\epsilon$ prescription used above is to ensure convergence in the $\lambda$ integrals. A separate $i\epsilon$ prescription is used to regulated the $t=0$ divergence. This is done by inserting a hard cutoff at $t=\cross$, and defining $\mathcal{C}$ by the following $i\epsilon$ prescription,
\begin{equation}\label{eq:ieps}
    \left(\int_{-\infty}^{-\cross}\frac{dt}{t}+\int_{\cross}^{\infty}\frac{dt}{t}\right)f(t)
    = \frac{1}{2}\lim_{\epsilon\rightarrow 0}\sum_{\pm}\int_{-i\infty\pm \epsilon}^{i\infty\pm \epsilon}\frac{dt}{t}f(-it) = \frac{1}{2}\int_{\mathcal{C}}\frac{dt}{t}f(-it)~.
\end{equation}
We note that this is the same $i\epsilon$ prescription used to regulate the de Sitter Wilson spool. The contour, $\mathcal{C}$, runs upward along the imaginary axis, as shown in the left panel of Figure~\ref{fig:Contour3}. With this regularization prescription, (\ref{eq:W}) becomes
\begin{equation}
    \log{D(\lambda)} = -\frac{i}{4}\int_{\mathcal{C}}\frac{dt}{t}\int_{\Gamma} z^2 \cot{(\pi z)}e^{it(z-\sqrt{1+\lambda})}dz~. 
\end{equation}
The contour integral over $\Gamma$ can be evaluated by summing the residues, where the poles occur at $z=n$ for $n\in\mathbb{N}$. This gives
\beq 
    \log{D(\lambda)}= \frac{i}{8}\int_{\mathcal{C}}\frac{dt}{t}\frac{\cos{(t/2)}}{\sin^3{(t/2)}}e^{-it\sqrt{1+\lambda}}~. \label{eq:sphereDl}\eeq
Using $D(\lambda)^{-1/2} = Z_\Delta$, we recognize the above is precisely the Wilson spool (\ref{eq:spoolds}) for the scalar field on $S^{3}$, 
where recall from \eqref{eq:lambdatoj} that the highest weight $j$ is related to $\lambda$ via $1+\sqrt{1+\lambda}=\Delta_+=-2j$: 
\begin{align} \label{eq:Wilsonfromtrace}
    \log{Z_\Delta} &= \frac{i}{4}\int_{\mathcal{C}}\frac{dt}{t}\frac{\cos{(t/2)}}{\sin{(t/2)}}\text{Tr}_{R_j}\left(\mathcal{P}e^{\frac{z}{2\pi}\oint a_L}\right)\text{Tr}_{R_j}\left(\mathcal{P}e^{-\frac{z}{2\pi}\oint a_R}\right)~.
\end{align}
While this is a familiar result from the Wilson spool literature~\cite{Castro:2023dxp,Castro:2023bvo}, we emphasize that it has been derived here purely from a Fredholm determinant-based trace formula for the three-sphere. Our derivation makes no use of Chern-Simons theory, up to the last step of identifying the exponential in \eqref{eq:sphereDl} as a trace over path ordered exponentials of the appropriate gauge connections.

It is interesting to note that our trace formula (equation \eqref{eq:trS3} with \eqref{eq:testfuncRp}) furnishes two different (but related) sets of geometric data. First, as in standard in trace formulae, the Poisson summation formula \eqref{eq:pois} for the three-sphere equates a sum over the eigenvalues $\lambda_\ell$ associated to the spectral operator \eqref{eq:Oform} to an integral over geodesic lengths $2\pi \abs{n}$ with $n\in \mathbb{Z}$ along the sphere. Further, equation \eqref{eq:trS3} shows that this geometric interpretation holds for any test function $h(z)$. Second, we have shown that when one specifies $h(z)$ as determined by the Fredholm determinant through \eqref{eq:resolvent}, the test function can be absorbed on the right hand (geometric) side of the trace formula to construct the holonomies in \eqref{eq:Wilsonfromtrace}. This converts the geometric data from particular geodesics along the sphere (with dependence on an unspecified test function) to an integral over topological invariants. We argue that the Fredholm determinant-based trace formula is of particular relevance, given that it precisely singles out this natural geometric data.



\section{Discussion} \label{sec:disc}

\noindent In this article, we revisited two particular methods for evaluating 1-loop determinants relevant for (A)dS$_{3}$ perturbative quantum gravity: the Selberg zeta function and the Wilson spool. The former relies on the group theoretic structure of the underlying spacetime, e.g., that the Euclidean BTZ black hole geometry is the quotient $\mathbb{H}^{3}/\mathbb{Z}$, while the latter relies heavily on the Chern-Simons formulation of 3D general relativity.

We uncovered two new insights. Firstly, for 1-loop determinants of arbitrary spin bosons on the Euclidean BTZ background, we showed that the (logarithm of) the Selberg zeta function is directly obtained through the Wilson spool.  In so doing, we found that the Selberg zeta function, historically framed as an Euler-product of primitive geodesics, can be recast as a product over descendants of highest or lowest weight states belonging to  the representation $\mathsf{R}_\Delta$ that are summed over in the Wilson spool. Equivalently, the Wilson spool, normally understood as a sum over Wilson loops that that wrap around a closed cycle, can be rephrased as a sum over primitive geodesics. 

Our second insight was a new derivation of the Wilson spool for a massive scalar on $S^{3}$. We achieved this by developing a trace formula for the trace of the resolvent associated with the scalar Laplacian on $S^{3}$. From this trace formula, we can directly construct the Wilson spool. Notably, the trace formula is not the standard spectral trace formula involving the trace of the heat kernel. Further, our derivation made no explicit reference of the topological nature of 3D general relativity. 

Our findings have a number of implications leading to new avenues worth exploring. 

\vspace{2mm}

\noindent \textbf{Spools beyond $S^{3}$.} In this work we focused on 1-loop determinants over the 3-sphere, $S^{3}$. According to the Gibbons-Hawking prescription, however, all classical saddle-point geometries should be included in the evaluation of the gravitational path integral (\ref{eq:gravpartgen}). On these grounds, our work captures only a single contribution to the Euclidean path integral. Notably, the infinite set of classical geometries for Euclidean dS$_{3}$ gravity can be explicitly enumerated and are known to be the quotients $S^{3}/\Gamma$, for $\Gamma$ a discrete freely acting subgroup of $SO(4)$ \cite{thurston1997three} (see also Appendix A of \cite{Castro:2011xb}). Among such discrete quotients are lens spaces $L(p,q)=S^{3}/\mathbb{Z}_{p}$, for coprime integers $p,q$. Such geometries were incorporated in the Euclidean dS$_{3}$ path integral in \cite{Castro:2011xb}, requiring an evaluation of the 1-loop determinants of kinetic operators of $S^{3}/\mathbb{Z}_{p}$. It remains an open question how to generalize the Wilson spool method for computing 1-loop determinants to lens space quotients, and $S^{3}/\Gamma$ more generally. 

One application of our work is that it can be used to suggest a candidate Wilson spool for lens spaces. All that is needed is an appropriate trace formula and a suitable test function, which can be obtained from the lens space Fredholm determinant. Indeed, as we derive in Appendix~\ref{app:lens}, the result from following our procedure is 

\begin{align}
    \mathbb{W}_j^{\text{lens}}
    &= -\frac{i}{p}\sum_{k\in\mathbb{Z}_p}\int_{\mathcal{C}}\frac{dt}{t}\frac{\cos{(t/2)}}{\sin{(t/2)}}\frac{\sin^2{(t/2)}\, e^{-it\sqrt{1+\lambda}}}{\left(\cos{(\frac{2\pi k}{p})-\cos{(t)}}\right)\left(\cos{(\frac{2\pi kq}{p})-\cos{(t)}}\right)}~,
\end{align}
where $1+\sqrt{1+\lambda}=\Delta_+=-2j$.
The integrand can be expressed in terms of non-standard characters, which can be taken off-shell. Note that this expression is equivalent to a similar character-based expression for lens space one-loop determinants, obtained by different methods in~\cite{Law:2021hwc}.

Likewise, our trace formula analysis in Appendix~\ref{app:higherd}  directly suggests a generalization for a Wilson spool in higher odd dimensions:
\begin{equation}
    \mathbb{W}_j^{\text{higher-d}}  = 2\left(\frac{i}{2}\right)^d\int_{\mathcal{C}}\frac{dt}{t}\frac{\cos{(t/2)}}{\sin^d{(t/2)}}e^{-it\sqrt{\frac{1}{4}(d-1)^2+\lambda}}~.
\end{equation}
As for the lens space case, similar expressions for one-loop determinants of higher-dimensional spheres written in terms of characters were obtained in~\cite{Anninos:2020hfj} (see also~\cite{Sun:2020ame} for higher-dimensional AdS); our contribution is to note that the correct form falls out naturally from our trace formula. This ``spool'' will no longer be topological, as this property is specific to three dimensions. It would be interesting to study in more detail how the structure of holonomies and non-standard representations lift to the higher dimensional case. In particular, it is crucial to understand better what Hilbert space is being traced over in the higher-dimensional case. One could also compare to non-topological definitions of Wilson loops in higher dimensions; see, for instance,~\cite{Costello:2013zra,Lacroix:2021iit}. 

\vspace{2mm}

\noindent \textbf{Descending to lower dimensions.} Specific theories of two-dimensional dilaton gravity, e.g., Euclidean (A)dS$_{2}$ Jackiw-Teteilboim (JT) gravity, serve as some of our best studied models of two-dimensional quantum gravity. A simplifying feature of JT gravity is that it can be cast as a topological BF theory. A prescription for the analog of the Wilson spool for JT gravity minimally coupled to massive matter fields was given in \cite{Fliss:2025sir}. It would be worth uncovering the connection between this lower-dimensional construction of the Wilson spool to the Selberg zeta function and trace formulae.

\vspace{2mm}

\noindent\textbf{Fermionic Fields.} So far the Wilson spool has only been constructed for scalars and higher spin bosonic fields. However, in the case of the Euclidean BTZ black hole, the Selberg zeta function formalism that has been related to the spool in this work naturally extends to fermionic fields as well \cite{Williams:2015azf,Keeler:2019wsx}. Thus, the analysis in Section \ref{2} provides a natural avenue to extend the Wilson spool construction to fermionic fields. We leave this for future work.

\vspace{2mm}

\noindent \textbf{Thermodynamics of near-extremal extremal black holes.} Methods for evaluating 1-loop determinants have important implications for the study of quantum corrections to the low-temperature thermodynamics of black holes \cite{Iliesiu:2020qvm,Iliesiu:2022onk,Turiaci:2023wrh,Banerjee:2023gll}. 1-loop quantum corrections at low-temperatures yield $\log(T)$ (for temperature $T$) modifications to the entropy, and give a gapless microscopic density of states for near-extremal non-supersymmetric black holes. For (near-horizon, near-extemal) Kerr-AdS$_{4}$ and rotating BTZ black holes, the $\log(T)$ correction comes from a $T^{3/2}$-scaling in the 1-loop partition
due to a careful regulation of zero modes using a small temperature expansion~\cite{Kapec:2023ruw,Rakic:2023vhv,Kolanowski:2024zrq,Kapec:2024zdj}. For near-extremal BTZ, this $T^{3/2}$ scaling can be demonstrated using the DHS method for computing 1-loop partition functions~\cite{Kapec:2024zdj}. In the course of~\cite{Kapec:2024zdj}, a precise identification of the branch of quasinormal modes responsible for the temperature scaling was uncovered. Black hole perturbations and quasi-normal modes have also featured in recovering the $T^{3/2}$ scaling for Kerr-(A)dS$_{4}$ black holes \cite{Arnaudo:2024bbd,Arnaudo:2025btb}. 

In the case of rotating BTZ, it would be interesting to re-examine these observations from the perspective of the Wilson spool, as it provides a version of the DHS method (with a natural regularization built in) for computing 1-loop determinants. Further, as we have seen, the trace formula provides a straightforward method to obtain DHS via the spool that can be generalized to both non-AdS spaces and to higher dimensions. It would be interesting to turn the crank, so to speak, to extend the quasinormal mode analysis to other near-extremal black holes, e.g., those with de Sitter asymptotics~\cite{Maulik:2025phe,Blacker:2025zca}. One caveat is that this would require careful consideration of the massless limit of the spool.

\subsection*{Acknowledgments}

\noindent It is a pleasure to thank Alejandra Castro, Jackson R. Fliss, Kurt Hinterbichler, Martin Huxley, Cynthia Keeler, Bob Knighton, Albert Law, Malcolm Perry and Rahul Poddar for helpful discussions. SH and CZ are supported by the National Science Foundation under Award Number 2412608. AS is supported by STFC grant ST/X000753/1, and is partially funded by the Royal Society under the grant ``Concrete Calculables in Quantum de Sitter.'' CZ also acknowledges participation in the Heising-Simons Foundation “Observational Signatures of Quantum Gravity” collaboration grants 2021-2818 and 2024-5305 as an affiliated member. 

\appendix

\section{Spectral functions for $S^3$} \label{app:genzetafuncs}

\noindent Let us briefly review spectral zeta functions corresponding to a general kinetic operator $\hat{\mathcal{O}}$ on the sphere:
\begin{equation}
    \hat{\mathcal{O}} =-\nabla^2 -\lambda\;.
\end{equation}
Here $\lambda$ is a real parameter, while
$-\nabla^2$ is the (massless) scalar Laplacian on the sphere $S^3$ with eigenvalues $\{\lambda_{\ell}\}$ degeneracies $d_{\ell}$,
\begin{equation}
    \lambda_\ell =\ell(\ell+2) \quad\text{and}\quad d_\ell=(\ell+1)^2\;,
\end{equation}
for $\ell\in\mathbb{Z}$. The  functional determinant $D(\lambda)$ of operator $\hat{\mathcal{O}}$ is given by
\begin{equation}\label{eq:detgen}
    D(\lambda) \equiv\text{det}(\hat{\mathcal{O}})=\prod_{\ell}(\lambda_{\ell}-\lambda)^{d_{\ell}}~,
\end{equation}
which in general will be an infinite product, and thus care is needed to ensure convergence. 

To evaluate the functional determinant (\ref{eq:detgen}), one often works with the (general) spectral zeta function 
\begin{equation}\label{eq:genzeta}
    \zeta_{\hat{\mathcal{O}}}(s,\lambda) \equiv \sum_{\ell=0}^{\infty}\frac{d_\ell}{(\lambda_\ell +\lambda)^s}~.
\end{equation}
For later convenience, we denote the zeta function for the special value $\lambda=0$ by $\zeta_{\hat{\mathcal{O}}}(s)\equiv \zeta_{\hat{\mathcal{O}}}(s,0)$. The zeta function (\ref{eq:genzeta}) may be expressed as a Mellin transform
\begin{equation}\label{eq:tr5}
    \zeta_{\hat{\mathcal{O}}}(s,\lambda) = \frac{1}{\Gamma(s)}\int_{0}^{\infty}\frac{dt}{t^{1-s}}\Theta(t)e^{-\lambda t}~,
\end{equation}
for ``partition function'' or heat kernel trace
\begin{equation}\label{eq:kernS} \Theta(t) = \sum_{\ell=0}^{\infty}d_\ell e^{-\lambda_\ell t}~.
\end{equation}
It is easy to verify that the logarithm of the functional determinant can be cast as
\begin{equation}\label{eq:tr6}
    \log{D(\lambda)} = -\lim_{s\rightarrow 0}\frac{\partial}{\partial s}\zeta_{\hat{\mathcal{O}}}(s,-\lambda) = -\lim_{s\rightarrow0}\frac{\partial}{\partial s}\biggl\{\frac{1}{\Gamma(s)}\int_{0}^{\infty}\frac{dt}{t^{1-s}}\Theta (t)e^{\lambda t}\biggr\}~.
\end{equation}

\subsection*{Functional and Fredholm determinant}

\noindent Let us now establish the relationship (\ref{eq:22}) between the $3$-sphere and lens space functional determinant $D(\lambda)$ and (regularized) Fredholm determinant 
\begin{equation}\label{eq:heatkern}
    \mathbf{\Delta} (\lambda) = \prod_{\ell=0}^{\infty}\left(1-\frac{\lambda}{\lambda_\ell}\right)^{d_\ell}e^{\frac{\lambda}{\lambda_\ell}d_\ell}\;.
\end{equation}
First, since the exponential in the expansion (\ref{eq:tr6}) is holomorphic everywhere on $\mathbb{C}$, we Taylor expand to find
\begin{align} \label{eq:logD}
    \log{D(\lambda)} = &-\lim_{s\rightarrow0}\frac{\partial}{\partial s}\biggl\{\frac{1}{\Gamma(s)}\int_{0}^{\infty}\frac{dt}{t^{1-s}}\Theta (t)e^{\lambda t}\biggr\} \nonumber \\
    = &-\lim_{s\rightarrow0}\frac{\partial}{\partial s}\biggl\{\frac{1}{\Gamma(s)}\int_{0}^{\infty}\frac{dt}{t^{1-s}}\Theta (t)\biggr\}-\lambda \lim_{s\rightarrow0}\frac{\partial}{\partial s}\biggl\{\frac{1}{\Gamma(s)}\int_{0}^{\infty}\frac{dt}{t^{-s}}\Theta (t)\biggr\} \nonumber \\ 
    &- \lim_{s\rightarrow0}\frac{\partial}{\partial s}\biggl\{\frac{1}{\Gamma(s)}\int_{0}^{\infty}\frac{dt}{t^{1-s}}\Theta (t)\sum_{m=2}^{\infty}\frac{\lambda^m}{m!}t^m\biggr\}~.
    \end{align}
By definition, the first term is
\begin{equation}\label{eq:fred3}
    \zeta_{\hat{\mathcal{O}}}'(0) \equiv \lim_{s\rightarrow0}\frac{\partial}{\partial s}\biggl\{\frac{1}{\Gamma(s)}\int_{0}^{\infty}\frac{dt}{t^{1-s}}\Theta (t)\biggr\}~.
\end{equation}
Inserting a hard cutoff to regulate the divergence at $t=0$, such that 
\be \lim_{s\rightarrow 0}\frac{1}{\Gamma(s)}\frac{\partial}{\partial s}\int_{\cross}^{\infty}\frac{dt}{t^{-s}}\Theta(t)= \lim_{s\rightarrow 0}\frac{1}{\Gamma (s)}\int_{\cross}^{\infty}dt t^{s}\log{(t)}\Theta(t)=0~,\ee
we see that the second term evaluates to
\begin{align}
    \lim_{s\rightarrow0}\frac{\partial}{\partial s}\biggl\{\frac{1}{\Gamma(s)}\int_{0}^{\infty}\frac{dt}{t^{-s}}\Theta (t)\biggr\} 
    &=\zeta_{\hat{\mathcal{O}}}(1)~.
\end{align}
The final term in (\ref{eq:logD}) may likewise be evaluated by removing the $t=0$ divergence using a hard cutoff, giving
\begin{align}\label{eq:fred2}
    \lim_{s\rightarrow0}\frac{\partial}{\partial s}\biggl\{\frac{1}{\Gamma(s)}\int_{\cross}^{\infty}\frac{dt}{t^{1-s}}\Theta (t)\sum_{m=2}^{\infty}\frac{\lambda^m}{m!}t^m\biggr\} 
    &= \sum_{m=2}^{\infty}\frac{\lambda^m}{m}\zeta_{\hat{\mathcal{O}}}(m)~.
\end{align}
It is easy to verify the result of (\ref{eq:fred2}) is equivalent to the logarithm of the Fredholm determinant,
\begin{align}
    \log{\mathbf{\Delta}(\lambda)} &= \log{\prod_{\ell=0}^{\infty}\left(1-\frac{\lambda}{\lambda_{\ell}}\right)^{d_\ell}}e^{\frac{\lambda}{\lambda_{\ell}}d_\ell} 
    = -\sum_{m=2}^{\infty}\frac{\lambda^m}{m}\zeta_{\hat{\mathcal{O}}}(m)~.
\end{align}
Altogether, the Fredholm determinant (\ref{eq:heatkern}) is related to the functional determinant by
\begin{equation}\label{eq:tr12}
    D(\lambda) =\mathbf{\Delta}(\lambda)e^{-\zeta_{\hat{\mathcal{O}}}'(0)}e^{-\lambda\zeta_{\hat{\mathcal{O}}}(1)}~,
\end{equation}
matching (\ref{eq:22}).

\section{Trace formulae and test functions beyond $S^3$}  \label{app:traceformbeyondS3}

\noindent Here we extend our analysis of Section \ref{sec:traceformS3} for spaces beyond the 3-sphere. In particular, below we will construct the appropriate trace formula and test function to give a proposal for the Wilson spool for a massive real scalar field on: (i) lens space quotients of $S^{3}$, and (ii) the higher-dimensional sphere $S^{d}$ for odd $d$.

\subsection{Lens spaces} \label{app:lens}

\noindent For a pair of coprime integers $(p,q)$ ($\text{gcd}(p,q)=1$), a lens space $L(p,q)$ is a quotient of the three-sphere by the discrete subgroup $\mathbb{Z}_{p}$ of the isometry group $SO(4)$ of $S^{3}$. It has Hopf geometry (\ref{Eq:desittermeteric}), but now with coordinate identifications
\beq (t_{E},\phi)\sim (t_{E},\phi)+\left(\frac{2\pi n}{p},\frac{2\pi n q}{p}+2\pi m\right)\;,\label{eq:Lpqid}\eeq
for integers $m,n$.\footnote{Note that the prime integer $q$ labels the various ways the cyclic group $\mathbb{Z}_{p}$ can be embedded into $SO(4)$. Further note that a shift in $q$ by an integer multiple of $p$ may be absorbed into an appropriate shift in integers $n,m$, and hence $q$ is defined up to mod $p$.} Such a coordinate identification makes the Euclidean metric (\ref{Eq:desittermeteric}) smooth at the horizon $\rho=\pi/2$. Note that the sphere itself is a special case of a lens space, $L(1,0)=S^{3}$.

The lens space identification (\ref{eq:Lpqid}) is generated by the operator $\rho\equiv e^{-2\pi\left(\frac{1}{p}H+i\frac{q}{p}J\right)}$,
where $H=i\partial_{t}$ is the generator of time-translations while $J\equiv i\partial_{\phi}$ the axial Killing vector. Thus, the lens space $L(p,q)$ can be thought of as defining a grand canonical ensemble with (inverse) temperature $\beta$ and angular potential $\vartheta$, 
\beq \beta=\frac{2\pi}{p}\;,\quad \vartheta=\frac{2\pi iq}{p}\;.\eeq
As such, a lens space is often referred to as a `thermal' quotient of the sphere.

\vspace{2mm} 

\noindent \textbf{Trace formula and test function.} The Fredholm determinant for the scalar Laplacian takes the same form as the sphere, $S^{3}$, given by (\ref{eq:fred}). 
The eigenspectrum is the same as the 3-sphere, $\lambda_\ell =\ell(\ell+2)$, but now the degeneracies are (cf. Eq. (43) of \cite{Martin:2020api} with $s=0$)
\begin{equation}\label{eq:ld}
    d_\ell = \frac{1}{p}\sum_{k\in\mathbb{Z}_p}\chi^F_{(\frac{\ell}{2})}(k\tau)\chi^F_{(\frac{\ell}{2})}(k\bar{\tau})\;,
\end{equation}
where $\chi^F_{(\ell)}(x)$ denote the standard (finite-dimensional) $\mathfrak{su}(2)$ characters,
\begin{equation}
    \chi^F_{(\ell)}(x) = \text{tr}_{(\ell)}(e^{\frac{ix}{2}\sigma_{3}}) = \frac{\sin{\left[\frac{x}{2}(2\ell+1)\right]}}{\sin{(x/2)}}~.
\end{equation}
 Here, $\sigma_{3}$ is the familiar Pauli matrix, $\ell$ is the highest weight of the representation, and $\tau$ and $\bar{\tau}$ are parameters defined as 
\begin{equation}
    \tau\equiv \frac{2\pi q}{p}-\frac{2\pi}{p}\;, \quad \bar{\tau} \equiv \frac{2\pi q}{p}+\frac{2\pi}{p}\;.
\end{equation}

These finite dimensional characters can also be expressed in terms of the non-standard characters \eqref{eq:chars} that appear in the de Sitter spool, using the relationship
\begin{equation}\label{eq:IDtoFDchars}
    \chi_{j}\left(\frac{x}{2\pi}\right)-\chi_{\bar{j}}\left(\frac{x}{2\pi}\right) = \frac{\sin{\left[\frac{x}{2}(2j+1)\right]}}{\sin{(\frac{x}{2})}}~.
\end{equation}
Recall also that
\begin{equation} \label{eq:Pinchars}
\chi_j\left(\frac{x}{2\pi}\right)\chi_{\bar{j}}\left(-\frac{x}{2\pi}\right) = -\sum_{k_1=0}^{\infty}\sum_{k_2=0}^{\infty}e^{ix(2j-k_1-k_2)}
=\frac{1}{4}\frac{e^{i x (2j+1)}}{\sin^2{(x/2)}}~.
\end{equation}

As before, the resolvent $R'(\lambda)$ is
\begin{align}\label{eq:Rplens}
    R'(\lambda) &\equiv -\frac{d^2}{d\lambda^2}\log{\mathbf{\Delta}(\lambda)}= \frac{1}{2p}\sum_{n\in\mathbb{Z}}\sum_{k\in\mathbb{Z}_p}\frac{d_k(n)}{(\lambda+1-n^2)^2}~,
\end{align}
where the degeneracy function is given by
\begin{equation}\label{eq:deglens}
    d_k(n) \equiv  \chi^F_{(\frac{1}{2}(n-1))}(k\tau)\chi^F_{(\frac{1}{2}(n-1))}(k\bar{\tau}) = \frac{\sin{\left[\frac{nk\pi}{p}(q-1)\right]}\sin{\left[\frac{nk\pi}{p}(q+1)\right]}}{\sin{\left[\frac{k\pi}{p}(q-1)\right]}\sin{\left[\frac{k\pi}{p}(q+1)\right]}}~,
\end{equation}
which is clearly even in $n$. 
The form of (\ref{eq:Rplens}) is well suited for applying the Poisson resummation formula since the test function is the same as that of $S^{3}$.  To arrive at our proposal for the Wilson spool, we will use the factorized form of the test function,
\begin{align}
    h(n)
    &= \frac{2\sqrt{1+\lambda}-n}{4\left(1+\lambda\right)^{3/2}(n-\sqrt{1+\lambda})^2}+\frac{2\sqrt{1+\lambda}+n}{4\left(1+\lambda\right)^{3/2}(n+\sqrt{1+\lambda})^2}~.
\end{align}
Applying the generic trace formula (\ref{eq:trS3}), we have
\begin{equation}
    \frac{d^2}{d\lambda^2}\log{D(\lambda)} = -\frac{i}{2p}\int_{\Gamma}\sum_{k\in\mathbb{Z}_p}d_k(z) h(z)\cot{(\pi z)} dz~.
\end{equation}
This demonstrates how straightforward it can be to generalize results using the trace formula.

\vspace{2mm}

\noindent \textbf{Functional determinant.} We can solve for the functional determinant $D(\lambda)$ by integrating twice with respect to $\lambda$ as done in the main text, see (\ref{eq:intd}). To safely interchange and evaluate the $\lambda$ integrals, we apply the Laplace transform given by (\ref{eq:lapl}), noting that $d_k(z)$ is even in $z$. We must also apply shift in contour before performing the $\lambda$ integrals as done in (\ref{eq:lamshift}). Doing so yields
\begin{align}\label{eq:Wlens}
    \log{D(\lambda)}  
    &= -\frac{i}{2p}\left[\int_{\cross}^{\infty}\frac{dt}{t}+\int_{-\infty}^{-\cross}\frac{dt}{t}\right]\int_{\Gamma} \sum_{k\in\mathbb{Z}_p}d_k(z)\cot{(\pi z)}e^{-t(z-\sqrt{1+\lambda})}dz~.
\end{align}
The $t=0$ divergence can be regulated using the $i\epsilon$ prescription given by (\ref{eq:ieps}), and the integral over $\Gamma$ can be evaluated by summing over the first order poles that occur at $z=n$ for $n\in\mathbb{Z}$, to obtain
\be \label{eq:simplelens}
    \log{D(\lambda)}=\frac{i}{2p}\sum_{k\in\mathbb{Z}_p}\int_{\mathcal{C}}\frac{dt}{t}\frac{\cos{(t/2)}}{\sin{(t/2)}}\frac{\sin^2{(t/2)}\, e^{-it\sqrt{1+\lambda}}}{\left(\cos{(\frac{2\pi k}{p})-\cos{t}}\right)\left(\cos{(\frac{2\pi kq}{p})-\cos{t}}\right)}~.
\ee
Note that the integrand can be expressed in terms of non-standard characters using \eqref{eq:IDtoFDchars},\eqref{eq:Pinchars} and the relation $\sqrt{1+\lambda}=-(2j+1)$. Using \eqref{eq:IDtoFDchars}, this can be re-expressed using the characters $\chi_j$ as
 \begin{align}
     \log{D(\lambda)} = \frac{i}{p}\sum_{k\in\mathbb{Z}}\int_{\mathcal{C}}\frac{dt}{t}\sin{(t)}\,\chi_{j}\left(\frac{t}{2\pi}-\frac{k}{p}\right)\chi_{j}\left(\frac{t}{2\pi}+\frac{k}{p}\right)  \chi_{j}\left(\frac{t}{2\pi}-\frac{kq}{p}\right)\chi_{j}\left(\frac{t}{2\pi}+\frac{kq}{p}\right) e^{it\sqrt{1+\lambda}}.
 \end{align}
We have left the factor $e^{it\sqrt{1+\lambda}}$ in the integrand, but note that this can also be expressed in terms of non-standard characters using \eqref{eq:Pinchars}. To recover the sphere result (\ref{eq:sphereDl}) from the above, simply set $k=0$.

We can also check that this matches known expressions for the lens space one-loop determinant. To do this, we perform the integral over $\Gamma$ but leave the sum over poles unevaluated, converting it to a sum over $l=n-1$. Using \eqref{eq:ld} and \eqref{eq:deglens}, the degeneracy functions $d_k(z)$ and the sum over $k$ can be re-expressed in terms of the original degeneracies, $d_\ell$. We find
\begin{equation}\label{eq:lenscheck}
    \log{D(\lambda)} = -\sum_{\ell=0}^{\infty}d_{\ell}\int_{\cross}^{\infty}\frac{dt}{t}e^{-t(\ell+1+\sqrt{\lambda+1})}-\sum_{\ell=0}^{\infty}d_{\ell}\int_{\cross}^{\infty}\frac{dt}{t}e^{-t(\ell+1-\sqrt{\lambda+1})}~.
\end{equation}
Recalling $\Delta_{\pm}\equiv 1\pm\sqrt{1+\lambda}$, 
and using the integral representation of logarithm function through the Schwinger parameter in \eqref{eq:Schwinger}, we see that the functional determinant (\ref{eq:lenscheck}) is
\begin{equation}
    \log{D(\lambda)} = \sum_{\pm}\sum_{\ell=0}^{\infty}d_{\ell}\log{(\ell+\Delta_{\pm})}~,
\end{equation}
consistent with, e.g., (42) of \cite{Martin:2020api}.

\subsection{Higher dimensional spheres} \label{app:higherd}

\noindent We can easily adapt our analysis to the odd $d$-dimensional sphere $S^{d}$. The (regularized) Fredholm determinant in this case is
\begin{equation}\label{eq:Sd}
    \mathbf{\Delta}(\lambda) = \prod_{\ell=0}^{\infty}\left(1-\frac{\lambda}{\lambda_\ell}\right)^{d_{\ell}}\text{exp}\left[d_{\ell}\sum_{p=1}^{\alpha}\frac{\lambda^p}{p\lambda_{\ell}^p}\right]~,
\end{equation}
where we note the order for $S^d$ is $\mu=\frac{d}{2}$, and we introduced $\alpha \equiv \left\lfloor \frac{d}{2} \right\rfloor$. The eigenvalues and degeneracies of the scalar Laplacian are
\begin{equation}
    \lambda_{\ell} = \ell(\ell+d-1) \quad \text{and} \quad d_{\ell}=\frac{(2\ell+d-1)(\ell+d-2)!}{\ell!(d-1)!}~.
\end{equation}
\par
Meanwhile, in higher dimensions the resolvent $R(\lambda)$ is defined in terms of higher order derivatives such that the exponential terms in (\ref{eq:Sd}) disappear. That is, 
\begin{align}\label{eq:resdim}
    R^{(\alpha)}(\lambda) &= -\frac{d^{\alpha+1}}{d\lambda^{\alpha+1}}\log{\mathbf{\Delta} (\lambda)}\nonumber \\ 
    &= \sum_{\ell=0}^{\infty}\frac{d_\ell \alpha!}{(\lambda_\ell-\lambda)^{\alpha+1}}~.
\end{align}
As before, the resolvent is related to the functional determinant, $D(\lambda)$.
It can be solved for by integrating the resolvent,
\begin{equation}
    \log{D(\lambda)} = - \prod_{p=1}^{\alpha+1}\int_{-\infty}^{\lambda}d\lambda_p R^{(\alpha)}(\lambda_p)~.
\end{equation}

Following the procedure outlined for $S^3$, we again factorize the Fredholm determinant. Defining
\begin{equation}
    \mathbf{\Delta}^{+}(z) = \prod_{\ell=0}^{\infty}\left(1-\frac{z}{\ell+\frac{d-1}{2}}\right)^{d_\ell}\text{exp}\left[d_{\ell}\sum_{p=1}^{\alpha+1}\frac{z^p}{p(\ell+\frac{d-1}{2})^p}\right]\;,
\end{equation}
we may factorize $\mathbf{\Delta}(\lambda)$ up to an overall entire function $Q(\ell)$,
\begin{align}\label{eq:fact}
    \mathbf{\Delta}(\lambda) &= Q(\ell)\prod_{\ell=0}^{\infty}\left(1-\frac{\kappa^2}{\left(\ell+\frac{d-1}{2}\right)^2}\right)^{d_\ell}\text{exp}\left[d_{\ell}\sum_{p=1}^{\alpha+1}\frac{\kappa^p+(-\kappa)^p}{p(\ell+\frac{d-1}{2})^p}\right]=Q(\ell)\mathbf{\Delta}^{+}(\kappa)\mathbf{\Delta}^{+}(-\kappa)~,
\end{align}
where
\begin{equation}
    \kappa^2= \left(\frac{d-1}{2}\right)^2+\lambda~.
\end{equation} 
\par
Consider, for example, $S^5$. Using (\ref{eq:fact}), the (log of the) Fredholm determinant can be written as a sum over $\mathbb{Z}$ as
\begin{align}\label{eq:fact2}
    \log&{\mathbf{\Delta}^{+}(\kappa)\mathbf{\Delta}^{+}(-\kappa)} = \sum_{\ell=0}^{\infty}d_\ell\log{\left[1+\frac{\kappa}{\ell+2}\right]}+\sum_{\ell=0}^{\infty}d_\ell\log{\left[1-\frac{\kappa}{\ell+2}\right]} +\sum_{\ell=0}^{\infty}\frac{d_\ell \kappa^2}{(\ell+2)^2}~
    \nonumber\\ 
    &= \frac{1}{12}\sum_{n=1}^{\infty}d(n)\log{\left[1+\frac{\kappa}{n+1}\right]} 
+\frac{1}{12}\sum_{n=3}^{\infty}d(-n)\log{\left[1+\frac{\kappa}{1-n}\right]} +\sum_{\ell=0}^{\infty}\frac{d_\ell \kappa^2}{(\ell+2)^2}~,
\end{align}
where $d(n)$ is the shifted degeneracy,
\begin{equation}
    d(n) =n(n+2)(n+1)^2~.
\end{equation}
For the first term in the second equality in \eqref{eq:fact2}, we shifted $n=\ell+1$, and we shifted $n=\ell+3$ for the second term. This is done to both convert into a sum over $\mathbb{Z}$, and to write the expression in a form such that parity in $n$ will be clear. We obtain
\begin{align}
    \log{\mathbf{\Delta}^{+}(\kappa)\mathbf{\Delta}^{+}(-\kappa)} &= \frac{1}{24}\sum_{n\in\mathbb{Z}}d(n)\log{\left[1+\frac{\kappa}{n+1}\right]}
+\frac{1}{24}\sum_{n\in\mathbb{Z}}d(-n)\log{\left[1+\frac{\kappa}{1-n}\right]} +\sum_{\ell=0}^{\infty}\frac{d_\ell \kappa^2}{(\ell+2)^2} ~.
\end{align}
From  (\ref{eq:resdim}), the resolvent for $S^5$ can be expressed as a sum over $\mathbb{Z}$ using the factorized Fredholm determinant (\ref{eq:fact2}). This can be written using the trace formula (\ref{eq:trS3}) with\footnote{The trace formula (\ref{eq:trS3}) can be used since the combination $d(n)h_{+}(n)+d(-n)h_{-}(n)$ is even in $n$.}
\begin{align}\label{eq:hdr}
    R^{(\alpha)}(\lambda)  &= \frac{1}{2}\sum_{n\in\mathbb{Z}}d(n)h_{+}(n) + \frac{1}{2}\sum_{n\in\mathbb{Z}}d(-n)h_{-}(n)\nonumber \\ 
    &= \frac{i}{2}\int_{\Gamma}d(z)h_{+}(z)\cot{(\pi z)}dz+\frac{i}{2}\int_{\Gamma}d(-z)h_{-}(z)\cot{(\pi z)}dz~,
\end{align}
where $h_{\pm}(n)$ is the test function associated with the factorized Fredholm determinant (\ref{eq:fact2}),
\begin{align}
    h_{\pm}(n) &= \frac{-35-8\lambda-9\sqrt{4+\lambda}\mp3n(2\pm n+3\sqrt{4+\lambda})}{96(4+\lambda)^{5/2}(1\pm n+\sqrt{4+\lambda})^3}~.
\end{align}
The resolvent can be integrated with respect to $\lambda$ to express the functional determinant for $S^5$ as
\begin{equation}
    \log{D(\lambda)} = -\frac{i}{2}\prod_{p=1}^{3}\int_{-\infty}^{\lambda_p}d\lambda_p \int_{\Gamma} [d(z)h_{+}(z)+d(-z)h_{-}(z)]\cot{(\pi z)}dz~.
\end{equation}
To allow the $\lambda_p$ integrals to be interchanged with the integral over $z$, we write the test function as a Laplace transformation,
\begin{align}
        h_{\pm}(z) = -\mathcal{L}\Bigl\{\frac{\pm3+3t\sqrt{4+\lambda}\pm t^2(4+\lambda)}{96(4+\lambda)^{5/2}}e^{\mp t(1+\sqrt{4+\lambda})}\Bigr\}(z)~.
\end{align}
The $\lambda_p$ integrals may be interchanged and evaluated by applying an $i\epsilon$ shift, as done in (\ref{eq:lamshift}), to ensure convergence. After the integrals are evaluated, we can take $\epsilon\rightarrow0$ to express the functional determinant as
\begin{align}
    \log{D(\lambda)} = -\frac{i}{24}\left[\int_{-\infty}^{-\cross}+\int_{\cross}^{\infty}\right]\frac{dt}{t}e^{t(1+\sqrt{4+\lambda})}\int_{\Gamma}d(z)\cot{(\pi z)}e^{-tz}dz~, 
\end{align}
where the cutoff at $z=\cross$ is to regulate the $t=0$ divergence. Lastly, we can the $i\epsilon$ prescription (\ref{eq:ieps}) used for $S^3$ to obtain:
\begin{align}
    \log{D(\lambda)} &= -\frac{i}{48}\int_{\mathcal{C}}\frac{dt}{t}e^{-it(1+\sqrt{4+\lambda})}\int_{\Gamma}d(z)\cot{(\pi z)}e^{itz} dz \nonumber\\  
    &= -\frac{i}{32}\int_{\mathcal{C}}\frac{dt}{t}\frac{\cos{(t/2)}}{\sin^5{(t/2)}}e^{-it\sqrt{4+\lambda}}~.
\end{align}
Note that although $d(z)$ is not itself even under $z \to -z$, the total sum of residues remains unchanged. This symmetry allows the contour integrals to be combined. 
\par
 This procedure can be applied for any odd dimension  sphere, $S^d$. The result for the functional determinant is
\begin{equation}
    \log{D(\lambda)} = \left(\frac{-i}{2}\right)^d\int_{\mathcal{C}}\frac{dt}{t}\frac{\cos{(t/2)}}{\sin^d{(t/2)}}e^{-it\sqrt{\frac{1}{4}(d-1)^2+\lambda}}~.
\end{equation}
Note the appearance of the familiar dimension-dependent phase \cite{Polchinski:1988ua}.

\bibliography{1looprefs}

\end{document}